\documentclass[aps,pre,twocolumn,notitlepage,superscriptaddress,showpacs, amsmath,  amssymb,nofootinbib]{revtex4-1}

\usepackage{color}
\usepackage{graphicx}
\usepackage{comment}
\usepackage[pagebackref=false]{hyperref}

\begin{document}
\title{Characterising information-theoretic storage and transfer in continuous time processes}
\author{Richard E. Spinney and Joseph T. Lizier}
\affiliation{Complex Systems Research Group \& Centre for Complex Systems, Faculty of Engineering and IT,
The University of Sydney, Sydney, New South Wales, Australia, 2006.
}
\date{\today}
\begin{abstract}
The characterisation of information processing is an important task in complex systems science. \emph{Information dynamics} is a quantitative methodology for modelling the intrinsic information processing conducted by a process represented as a time series, but to date has only been formulated in discrete time. Building on previous work \cite{spinney_transfer_2017} which demonstrated how to formulate transfer entropy in continuous time, we give a total account of information processing in this setting, incorporating information storage. We find that a convergent rate of predictive capacity, comprised of the transfer entropy and active information storage, does not exist, arising through divergent rates of active information storage. We identify that active information storage can be decomposed into two separate quantities that characterise predictive capacity stored in a process: active memory utilisation and instantaneous predictive capacity. The latter involves prediction related to path regularity and so solely inherits the divergent properties of the active information storage, whilst the former permits definitions of pathwise and rate quantities.
We formulate measures of memory utilisation for jump and neural spiking processes and illustrate measures of information processing in synthetic neural spiking models and coupled Ornstein-Uhlenbeck models.
The application to synthetic neural spiking models demonstrates that active memory utilisation for point processes consists of discontinuous jump contributions (at spikes) interrupting a continuously varying contribution (relating to waiting times between spikes), complementing the behaviour previously demonstrated for transfer entropy in these processes.
\end{abstract}
\maketitle
\section{Introduction}
Information dynamics \cite{lizier_local_2008,lizier_measuring_2014,lizier_information_2010,lizier_local_2012} is a framework that seeks to characterise distributed computation by identifying primitives of information processing in autonomously evolving or `computing' systems.
Specifically, this involves modelling how information is stored in and transferred between variables in the system when the values of these variables are dynamically updated in time.
These primitive information processing operations studied by the framework, information storage and transfer, are measured by the active information storage \cite{lizier_local_2012} and transfer entropy \cite{schreiber_measuring_2000} (and higher orders thereof in larger multivariate systems \cite{lizier_information_2010,lizier_local_2013}) respectively.
The measures in this framework have been used to characterise and explain behaviour observed in various complex systems, providing novel insights across a broad range of fields including in: canonical complex systems such as cellular automata \cite{lizier_local_2013} and random Boolean networks \cite{lizier_smallworld_2011}, interpretation of dynamics in \cite{williams_information_2010} and improved algorithms for machine learning \cite{dasgupta_information_2013}, characterising information processing signatures in biological signalling networks \cite{walker_informational_2016}, collective behaviour in swarms \cite{tom16a,wang_quantifying_2012}, non-linear time-series forecasting \cite{gar16a}, and in computational neuroscience applications in identifying neural information flows from brain imaging data \cite{faes_conditional_2014,timme_multiplex_2014,ito_extending_2011}, inferring effective network structure \cite{faes15b,wibral_transfer_2011,vicente_transfer_2011}, providing evidence for the predictive coding hypothesis \cite{brodski_2017} and identifying differences in predictive information in autism spectrum disorder subjects and controls \cite{gomez_2014}.
\par
However, to date, no complete theoretical account for how this framework should be applied to continuous time systems have been offered, despite such systems being ubiquitous in fields throughout all of science. Recently we have given an account of how to formulate transfer entropy in such systems \cite{spinney_transfer_2017}. In this paper we build upon these developments by addressing the concept of predictive capacity, comprising the transfer entropy and active information storage, in these systems. We find that such a quantity is much more complicated owing to the predictive properties that can be derived from the regularity properties of continuous time processes. As such, we focus in this paper on investigating the nature of the active information storage in continuous time processes. To proceed we find that it is necessary to decompose the active information storage into two components related to two distinct characteristics of information processing. We call these two quantities \emph{active memory utilisation} and \emph{instantaneous predictive capacity}. The former is designed to behave as a rate and to be complementary to transfer entropy, whilst the latter relates intrinsic uncertainty and path regularity properties of the process and may be understood asymptotically and independently to both transfer entropy and active memory utilisation. 
The results bring our understanding of the nature of information storage in such systems in line with that of information transfer, which is important not only in that they are both fundamental components of models of information processing, but also because of the insight these results bring to the growing role of information storage analyses in neural imaging data in particular \cite{gomez_2014,brodski_2017,faes_conditional_2014}.
Our insights reveal, for example, the expected behaviour of active information storage and its components when measured on discrete-time samples of underlying continuous-time processes, whereby the active memory utilisation is the only quantity that will approach a limiting value as the discrete time step approaches zero; this has significant implications for empirical analyses.
\par
The overall plan of the paper is as follows. We shall introduce information dynamics as currently conceptualised. 
Following on from this we outline the central problem under consideration, the definition and identification of predictive information associated with information storage in continuous time. 
Next we present a set of postulates from which we deduce the central division of the active information storage, identifying the active memory utilisation and instantaneous predictive capacity, the behaviour of which, and their relation to other information theoretic measures of stochastic processes, is then detailed. 
Such quantities are calculated for some pertinent applications, including simple neural spiking models, where an understanding of information processing in continuous time is a pressing concern. Finally, a full information processing description, consisting of both active memory utilisation and transfer entropy, is provided for a more complicated model of neural spiking with two spike trains and a coupled Ornstein-Uhlenbeck process, where these examples were selected such that the results may complement previous demonstrations of the properties of the transfer entropy in continuous time \cite{spinney_transfer_2017,barn17a}.
\section{Information Dynamics}
First, we summarise the basic principles of information dynamics as formulated in discrete time.
Central to the concept of information dynamics is the idea that each evolving state can be modelled as being `computed' from the past, in the sense of an `intrinsic computation' \cite{feldman_organization_2008}, and that this computation is characterised by the \emph{predictive capacity}. This concept is made concrete in the context of two random processes $X_{\{0:m\}}=\{X_0,\ldots,X_m\}$ and $Y_{\{0:m\}}=\{Y_0,\ldots,Y_m\}$ taking individual outcomes $x_{\{0:m\}}=\{x_0,\ldots,x_m\}$ and $y_{\{0:m\}}=\{y_0,\ldots,y_m\}$, wherein the predictive capacity of the state $X_{n+1}$ at time $n< m$ is considered, axiomatically, to be
\begin{quote}
the reduction of uncertainty in $X_{n+1}$ that arises from knowing the path histories $X_{\{0:n\}}=x_{\{0:n\}}$ and $Y_{\{0:n\}}=y_{\{0:n\}}$, at time $n$, over having no other knowledge.
\end{quote}
 This predictive capacity is then formalised mathematically as $C_X$, given as a mutual information or the difference between two (conditional) entropies \cite{lizier_information_2010,lizier_local_2013}
\begin{align}
C_X&=I(X_{n+1};X_{\{0:n\}},Y_{\{0:n\}})\nonumber\\
&=H(X_{n+1})-H(X_{n+1}|X_{\{0:n\}},Y_{\{0:n\}})
\end{align}
based on underlying ensemble probabilities, $p$. Following on from this quantification of predictive capacity, the central approach and framework of information dynamics is to decompose such a quantity into two specific terms with this decomposition termed the \emph{computational signature}, viz.
\begin{align}
C_X&=H(X_{n+1})-H(X_{n+1}|X_{\{0:n\}})\nonumber\\
&\quad+H(X_{n+1}|X_{\{0:n\}})-H(X_{n+1}|X_{\{0:n\}},Y_{\{0:n\}})\nonumber\\
&=I(X_{n+1};X_{\{0:n\}})+I(X_{n+1};Y_{\{0:n\}}|X_{\{0:n\}})\nonumber\\
&=A_X+T_{Y\to X}
\end{align}
where $A_X$ is known as the \emph{active information storage} \cite{lizier_local_2012} and is explicitly written
\begin{align}
A_X&=H(X_{n+1})-H(X_{n+1}|X_{\{0:n\}})\nonumber\\
&=\mathbb{E}\left[ \ln\frac{p(x_{n+1}|x_{\{0:n\}})}{p(x_{n+1})}\right],
\end{align}
and where $T_{Y\to X}$ is well known in many distinct areas of science as the \emph{transfer entropy} \cite{schreiber_measuring_2000,kaiser_information_2002,barnett_transfer_2012,bossomaier_introduction_2016} and is explicitly written
\begin{align}
T_{Y\to X}&=H(X_{n+1}|X_{\{0:n\}})-H(X_{n+1}|X_{\{0:n\}},Y_{\{0:n\}})\nonumber\\
&=\mathbb{E}\left[ \ln \frac{p(x_{n+1}|x_{\{0:n\}},y_{\{0:n\}})}{p(x_{n+1}|x_{\{0:n\}})} \right ].
\end{align}
Here we note the shorthand $p(x_{n+1}|x_{\{0:n\}},y_{\{0:n\}})=p(X_{n+1}=x_{n+1}|X_{\{0:n\}}=x_{\{0:n\}},Y_{\{0:n\}}=y_{\{0:n\}})$ etc. which we use to indicate, where appropriate, that the arguments of the probabilities are simply specific realisations of the variables to which the probabilities correspond, and the notation $\mathbb{E}[\ldots]$ which denotes an ensemble average with respect to $p(x_{\{0:n+1\}},y_{\{0:n\}})$. Since both contributions can be written as (conditional) mutual informations both are rigorously non-negative. 
\par
The computational signature effects a model of intrinsic computation based on this partition into $A_X$ and $T_{Y\to X}$, such that there exists an identifiable storage component attributed to the past of X through $A_X$, plus an information transfer component attributed to the past of Y in the context of X through $T_{Y\to X}$ \cite{lizier_local_2013}.
\par
More recent developments have emphasised that information theoretic quantities should be constructed from suitable \emph{local} or \emph{pointwise} quantities, of which the ensemble quantities, $A_X$, $T_{Y\to X}$ etc. are suitable expectations. Consequently we recognise the structure of the \emph{local active information storage}, $a_X$, given by $A_X=\mathbb{E}[ a_x(x_{n+1},x_{\{0:n\}})]$ \cite{lizier_local_2012} and \emph{local transfer entropy}, $t_{Y\to X}$, given by $T_{Y\to X}=\mathbb{E}[ t_{Y\to X}(x_{n+1},x_{\{0:n\}},y_{\{0:n\}})]$ \cite{lizier_local_2008}. Explicitly, we have
\begin{align}
a_X(x_{n+1},x_{\{0:n\}})&=\ln\frac{p(x_{n+1}|x_{\{0:n\}})}{p(x_{n+1})}\nonumber\\
t_{Y\to X}(x_{n+1},x_{\{0:n\}},y_{\{0:n\}})&=\ln\frac{p(x_{n+1}|x_{\{0:n\}},y_{\{0:n\}})}{p(x_{n+1}|x_{\{0:n\}})}.
\end{align}
We then also define the total computational signature, on a local scale, $c_X=a_X+t_{Y\to X}$. It is important to note that such local values have no bound on their sign and thus may be negative. Such an approach allows significance to placed on \emph{single realisations} of a process, allowing fine characterisation of spatial temporal features, such as the identification of dynamics that are informative, but especially those which are \emph{misinformative}, characterised by negative local values (which have been shown to identify interesting aspects of dynamics in cellular automata \cite{lizier_local_2008,lizier_local_2012} and stimulus changes in the cat visual cortex \cite{wibral_2014}).
\section{Formulating quantities in continuous time}
\subsection{Background and established quantities}
As originally formulated, information dynamics treats only systems in discrete time. However, many systems of great interest, not only to complexity research, but many areas of science are naturally formulated in continuous time. In such systems time series are not indexed by the integers, but by connected subsets of the real line such that instead of collections of random variables, one deals with random functions for which we use notation $X_{\mathcal{A}}=\{X(t'):t'\in \mathcal{A}\}$ with individual realisations $x_{\mathcal{A}}$. In previous work \cite{spinney_transfer_2017} we established how to treat the transfer entropy in continuous time. Important consequences of this work were the recognition that one must consider a transfer entropy \emph{rate}, and that this is formulated from an expectation of a \emph{functional} that assigns a number to individual realisations of the process called the \emph{pathwise transfer entropy}, $\mathcal{T}^{[t_0,t]}_{Y\to X}[x_{[\tau,t]},y_{[\tau,t)}]$, which represents the total accumulated predictive capacity transferred from $Y$ in the context of the history of $X$, on the interval $[t_0,t]$ as a function of the path realisations  $x_{[\tau,t]}$ and $y_{[\tau,t)}$, with $\tau\leq t_0<t$. Importantly, these are constructed from probability measures on \emph{complete paths}, emphasising that quantifications of evolving sequences should make sense quantitatively and conceptually in the wider context of them being understood, rigorously, as outcomes of fully realised stochastic processes. We may summarise this through the expressions for the transfer entropy rate and its relation to the pathwise transfer entropy, for a process with a defined time origin at $\tau$,
\begin{align}
T^{[t_0,t]}_{Y\to X}&=\mathbb{E}[\mathcal{T}^{[t_0,t]}_{Y\to X}[x_{[\tau,t]},y_{[\tau,t)}]]\nonumber\\
&=\int_{t_0}^t\dot{T}_{Y\to X}(t')dt'\nonumber\\
\dot{T}_{Y\to X}(t)&=\frac{d}{dt}\mathbb{E}[ \mathcal{T}^{[t_0,t]}_{Y\to X}[x_{[\tau,t]},y_{[\tau,t)}]]\nonumber\\
\mathcal{T}^{[t_0,t]}_{Y\to X}[x_{[\tau,t]},y_{[\tau,t)}]&=\ln\frac{d\mathbb{P}_{X|\{Y\}}[x_{(t_0,t]}|x_{[\tau,t_0]},\{y_{[\tau,t)}\}]}{d\mathbb{P}_{X}[x_{(t_0,t]}|x_{[\tau,t_0]}]}
\label{Trate}
\end{align}
where $T^{[t_0,t]}_{Y\to X}$ is the expected, cumulative, `transferred' information on the interval $[t_0,t]$ equal to the expected pathwise transfer entropy or the integral of the transfer entropy rate, $\dot{T}_{Y\to X}(t)$, over the interval. This property arises from the pathwise transfer entropy explicitly being the logarithm of a Radon-Nikodym (RN) derivative between probability measures on $x_{(t_0,t]}$. The central quantity, the pathwise transfer entropy, exists when such a RN derivative exists and the measures are absolutely continuous with respect to each other. 
 One may, non-rigorously, but safely in most instances, consider such a quantity to be the ratio of path `probabilities' defined through the following
\begin{align}
&\frac{d\mathbb{P}_{X|\{Y\}}[x_{(t_0,t]}|x_{[\tau,t_0]},\{y_{[\tau,t)}\}]}{d\mathbb{P}_X[x_{(t_0,t]}|x_{[\tau,t_0]}]}
\sim\nonumber\\
&\qquad\quad\lim_{n\to\infty}\prod_{i=0}^n\frac{p(x_{i+1}|x_{\{-k:i\}},y_{\{-k:i\}})}{p(x_{i+1}|x_{\{-k:i\}})},
\end{align}
where $t_0=0$, $x_i\equiv x_{i\Delta t}$ and $\Delta t=t/(n+1)=-\tau/k$. We note the construction of $\mathbb{P}_{X|\{Y\}}[x_{(t_0,t]}|x_{[\tau,t_0]},\{y_{[\tau,t)}\}]\sim \prod_{i=0}^np(x_{i+1}|x_{\{-k:i\}},y_{\{-k:i\}})$, emphasising that this asymptotic product form in general is not equal to the analogously constructed usual conditional probability $p(x_{\{1:n+1\}}|x_{\{-k:0\}},y_{\{-k:n\}})$.
\par
When dealing with integrated quantities such as the pathwise transfer entropy, which represents the accumulated transfer of information on the interval, one may construct a rate in an alternative sense viz.
\begin{align}
\mathring{T}_{Y\to X}&=\lim_{t\to\infty}\frac{1}{t-t_0}\mathbb{E}\left[\ln\frac{d\mathbb{P}_{X|\{Y\}}[x_{(t_0,t]}|x_{[\tau,t_0]},\{y_{[\tau,t]}\}]}{d\mathbb{P}_{X}[x_{(t_0,t]}|x_{[\tau,t_0]}]}\right].
 \end{align}
 When the process is stationary we have $\mathring{T}_{Y\to X}=\dot{T}_{Y\to X}$. Moreover, in addition, when the process is ergodic it may be expressed
 \begin{align}
\mathring{T}_{Y\to X}&=\lim_{t\to\infty}\frac{1}{t-t_0}\ln\frac{d\mathbb{P}_{X|\{Y\}}[x_{(t_0,t]}|x_{[\tau,t_0]},\{y_{\tau,t)}\}]}{d\mathbb{P}_{X}[x_{(t_0,t]}|x_{[\tau,t_0]}]}
 \end{align}
 forming the basis for any empirical estimation methods.
\subsection{Extension to active information storage and issues}
The primary motivation of this paper is to extend the full description of information processing in intrinsic computation, outlined in the previous section, to the case of continuous time. To do so in keeping with the above definitions of the transfer entropy rate is straightforward: the predictive capacity and subsequent computational signature is defined over some time interval $\Delta t$, with mean rates emerging when dividing by $\Delta t$ as the limit $\Delta t\to 0$ is taken. Firstly, the local predictive capacity, before the notion of a rate introduced, is therefore captured by
\begin{align}
c_X&=\lim_{dt\to 0}\ln\frac{p(x_{t+dt}|x_{[\tau,t]},y_{[\tau,t]})}{p(x_{t+dt})}\nonumber\\
&=\lim_{dt\to 0}\ln\frac{p(x_{t+dt}|x_{[\tau,t]})}{p(x_{t+dt})}\nonumber\\
&\quad+\ln\frac{p(x_{t+dt}|x_{[\tau,t]},y_{[\tau,t]})}{p(x_{t+dt}|x_{[\tau,t]})}\nonumber\\
&=a_X+t_{Y\to X}.
\end{align}
A \emph{rate} of (average) predictive capacity would be then constructed as follows
\begin{align}
\dot{C}_X&=\frac{d}{dt}\mathbb{E}[{c}_X]\nonumber\\
&=\lim_{dt\to 0}\frac{1}{dt}\mathbb{E}\left[\ln\frac{p(x_{t+dt}|x_{[\tau,t]},y_{[\tau,t]})}{p(x_{t+dt})}\right]\nonumber\\
&=\lim_{dt\to 0}\frac{1}{dt}\mathbb{E}\left[\ln\frac{p(x_{t+dt}|x_{[\tau,t]})}{p(x_{t+dt})}\right]\nonumber\\
&\quad+\frac{1}{dt}\mathbb{E}\left[\ln\frac{p(x_{t+dt}|x_{[\tau,t]},y_{[\tau,t]})}{p(x_{t+dt}|x_{[\tau,t]})}\right]\nonumber\\
&=\frac{d}{dt}\mathbb{E}[{a}_X]+\frac{d}{dt}\mathbb{E}[{t}_{Y\to X}]
\end{align}
where $d\mathbb{E}[{t}_{Y\to X}]/dt$ emerges as the previously defined transfer entropy rate $\dot{T}_{Y\to X}$ and $d\mathbb{E}[{a}_{X}]/dt$ is an analogously defined `active information storage rate'. However, there are significant issues associated with this quantity, $d\mathbb{E}[ {c}_X]/dt$, stemming from the proposed contribution, $d\mathbb{E}[{a}_{X}]/dt$.
\par
To understand the problem with such a quantity it is helpful to reconstruct it in the manner of the definition of the transfer entropy rate in Eq.~(\ref{Trate}) through the definition of a hypothetical `pathwise active information storage'. $\mathcal{A}^{[t_0,t]}_X[x_{[\tau,t]}]$, viz.
\begin{align}
\mathcal{A}^{[t_0,t]}_{X}[x_{[\tau,t]}]&=\ln\frac{d\mathbb{P}_X[x_{(t_0,t]}|x_{[\tau,t_0]}]}{d\mathbb{P}^{\emptyset}_{X}[x_{(t_0,t]}]}
\label{Arate}
\end{align}
with
\begin{align}
&\frac{d\mathbb{P}_X[x_{(t_0,t]}|x_{[\tau,t_0]}]}{d\mathbb{P}^{\emptyset}_{X}[x_{(t_0,t]}]}
\sim\lim_{n\to\infty}\prod_{i=0}^n\frac{p(x_{i+1}|x_{\{-k:i\}})}{p(x_{i+1})}.
\end{align}
Here the issue emerges: this quantity does not converge since the limit of the denominator does not lead to a measure on $x_{(t_0,t]}$, i.e. $\mathbb{P}^{\emptyset}_{X}$ does not exist. It then follows that a finite active information storage rate, $d\mathbb{E}[a_X]/dt$, does not exist either. It is instructive to demonstrate this, and the general problem, with a brief example.
\par
To demonstrate the issue with identifying predictive capacities and active information storage as currently defined in continuous time we consider a simple Ornstein Uhlenbeck process in stochastic differential equation form
\begin{align}
dx_t&=-\kappa x_tdt+\sigma dW_t,
\label{OU}
\end{align}
where $W_t$ is a Wiener process. Since there is no extrinsic process (i.e. $Y$) to consider, the total predictive capacity of the natural computation is identical to the active information storage. When formulating such a prediction we may leave the time horizon over which such a prediction is made to be a free parameter, $\Delta t$. Doing so leads to a parametric predictive capacity/active information storage, $C_X^{(\Delta t)}=A_X^{(\Delta t)}$ given by
\begin{align}
A_X^{(\Delta t)}&=\lim_{(t-\tau)\to\infty}\mathbb{E}\left[ \ln\frac{p(x_{t+\Delta t}|x_{[\tau,t]})}{p(x_{t+\Delta t})}\right]\nonumber\\
&=\mathbb{E}\left[ \ln\frac{p(x_{t+\Delta t}|x_t)}{p(x_{t+\Delta t})}\right]\nonumber\\
&=\frac{1}{2}\ln\left[\frac{e^{\kappa\Delta t}}{e^{\kappa\Delta t}-e^{-\kappa\Delta t}}\right],
\label{AISOU}
\end{align}
which is easily calculated using the well known transition probability density of the Ornstein-Uhlenbeck process \cite{risken_fokker-planck_1996}, detailed in Appendix \ref{appA01}, where it has been assumed that the process is in the stationary state. If we attempt to construct a rate in the limit $\Delta t\to 0$, it diverges. Moreover, even an attempt to find an $\mathcal{O}(1)$ quantity in the limit $\Delta t\to 0$ produces a divergent quantity. This has a simple interpretation: with knowledge of the process' history, the sampling paths of the process allow for arbitrary precision in the prediction at smaller and smaller time horizons. This is reflected in the numerator of $a_X$ which tends to a delta function. However, this contribution is simply unmatched by the uncertainty without conditioning on the process' history appearing in the denominator, which, as a Shannon (differential) entropy, remains bounded independently of the prediction horizon.
\par
In this sense we see that the active information storage is actually performing precisely as it should: one \emph{can} be infinitely more precise over an infinitesimal time horizon with knowledge of the process' history than with an isolated prediction for such processes. This motivates many additional questions: is such a measure of stored predictive information appropriate? Can it be decomposed into divergent and non-divergent terms in the $\Delta t\to 0$ limit and do these quantities possess sensible interpretations that can be meaningfully explored? We take the position that the active information storage decomposes into two distinct quantities related to active memory utilisation and instantaneous predictive capacity. These are detailed in the subsequent sections.

\section{Decomposition of stored information into active memory utilisation and instantaneous predictive capacity}
Here we posit that in general the active information storage $A_X$ describes a generalised sense of memory utilisation and is, in fact, comprised of two quantities. The first is related to memory, understood in an intuitive manner, whilst the second does not characterise memory, but the predictive capacity that is obtained solely from the current state of the system we term `instantaneous prediction'. We will demonstrate that these, in general, describe two distinct features of stochastic processes. The quantity that describes memory is a dynamical quantity that possesses a rate which we call the \emph{active memory utilisation rate}, $\dot{M}_X$. The instantaneous prediction is a non-dynamical quantity not amenable to description as a rate which we call the \emph{instantaneous predictive capacity}, $I_X$, where
\begin{align}
A_X&=I_X + \dot{M}_Xdt+\mathcal{O}(dt^2),
\end{align}
with $I_X \geq 0$ and $\dot{M}_X\geq 0$. We point out that $\dot{M}dt$ need not, however, comprise all $\mathcal{O}(dt)$ contributions in $A_X$, since $I_X$ may have $\mathcal{O}(dt)$ components also.
 We arrive at such a division through the introduction of the following postulates designed to construct a quantity $\dot{M}_X$ which is complementary to $\dot{T}_{Y\to X}$:
\begin{enumerate}
\item Measures of memory utilisation should assign finite, unit-less, values to complete path realisations of a time series/stochastic process.
\item Measures of memory utilisation should be formed from RN derivatives between equivalent probability measures\footnote{Equivalent probability measures are those which are absolutely continuous with respect to each other such that they agree on which sets of events have zero probability. Informally, this is to be understood as them agreeing on which realisations are `possible'.} on path realisations.
\item The active information storage \emph{contains} memory utilisation such that any decomposition yields positive quantities in expectation.
\item The memory utilisation is found by \emph{maximsing} such a component of active information storage such that the first two postulates are met.
\end{enumerate}
Informally, this is to be interpreted as the requirement that, regardless of time basis, we should i) be able to discuss the memory that has been cumulatively utilised over finite intervals ii) measure memory by comparing the relative weight assigned to the paths over these intervals by two models of the behaviour which agree on which paths are possible and that iii) find the largest mean contribution that achieves this with the currently existing measure of information storage which is derived from the, axiomatically fundamental, predictive capacity.
\par
In order to meet the second and third postulates, we must decompose $A_X$ through the introduction of a new transition probability, $\mathcal{P}$, that converges to a measure which is equivalent to $p[x_{[t_0,t]}]$, but also describes the statistics of the process such that it is an ensemble probability, $p$, itself. This ensures that it is a component of $A_X$, ensuring both $I_X\geq 0$ and $\dot{M}_X\geq 0$. To achieve the first two postulates we must retain the path regularity of the process and thus include $x_t$ in the condition of the transition probability, along with an unspecified additional component $\mathcal{A}\subseteq x_{[\tau,t)}$ required for positivity related to the third postulate. To achieve this we write
\begin{align}
A_X &= \lim_{dt\to 0}\mathbb{E}\left[\ln\frac{p(x_{t+dt}|x_{[\tau,t]})}{p(x_{t+dt})}\right]\nonumber\\
&= \lim_{dt\to 0}\mathbb{E}\left[\ln\frac{\mathcal{P}(x_{t+dt}|x_{[\tau,t]})}{p(x_{t+dt})}\right]+\mathbb{E}\left[\ln\frac{p(x_{t+dt}|x_{[\tau,t]})}{\mathcal{P}(x_{t+dt}|x_{[\tau,t]})}\right]\nonumber\\
&= \lim_{dt\to 0}\mathbb{E}\left[\ln\frac{p(x_{t+dt}|x_{t}\cup\{\mathcal{A}\subseteq x_{[\tau,t)}\})}{p(x_{t+dt})}\right]\nonumber\\
&\quad+\mathbb{E}\left[\ln\frac{p(x_{t+dt}|x_{[\tau,t]})}{p(x_{t+dt}|x_{t}\cup\{\mathcal{A}\subseteq x_{[\tau,t)}\})}\right]\nonumber\\
&=I_X + \dot{M}_Xdt.
\end{align}
The portion explainable as a dynamic memory source, $\dot{M}_X$, is then separated from the remainder $I_X$ by maximising $\dot{M}_X$ with respect to $\mathcal{A}$ such that we have
\begin{align}
\dot{M}_X&=\max_{\mathcal{A}}\lim_{dt\to 0}\frac{1}{dt}\mathbb{E}\left[\ln\frac{p(x_{t+dt}|x_{[\tau,t]})}{p(x_{t+dt}|x_{t}\cup\{\mathcal{A}\subseteq x_{[\tau,t)}\})}\right]\nonumber\\
&=\lim_{dt\to 0}\frac{1}{dt}\mathbb{E}\left[\ln\frac{p(x_{t+dt}|x_{[\tau,t]})}{p(x_{t+dt}|x_{t})}\right]
\label{Mdot}
\end{align}
corresponding to $\mathcal{A}=\emptyset$ and $\mathcal{P}(x_{t+dt}|x_{[\tau,t]})=p(x_{t+dt}|x_t)$ due to the properties of conditioning in mutual informations.  
\par
 This then defines the active memory utilisation rate, leaving the instantaneous predictive capacity to be given by
\begin{align}
I_X&=\lim_{dt \to 0}\mathbb{E}\left[\ln\frac{p(x_{t+dt}|x_{t})}{p(x_{t+dt})}\right].
\end{align}
Explicitly, the active memory utilisation vanishes for Markov, i.e. \emph{memoryless}, processes. Meanwhile, the instantaneous predictive capacity, whilst not permitting a rate since the numerator encodes path regularity whilst the denominator does not, will lie in $[0,\infty]$ due to its form as a mutual information.
\subsection{Active memory utilisation in continuous time}
Here we describe in more detail the behaviour and form of the active memory utilisation in continuous time which very closely follows the form of the transfer entropy \cite{spinney_transfer_2017}. As with the transfer entropy, whilst Eq.~(\ref{Mdot}) describes the mean rate of active memory utilisation, a local, or pointwise rate,
\begin{align}
\dot{m}_X[t,x_{[\tau,t]}]&=\lim_{dt\to 0}\frac{1}{dt}\ln\frac{p(x_{t+dt}|x_{[\tau,t]})}{p(x_{t+dt}|x_{t})}
\label{mlocrate}
\end{align}
is not guaranteed to exist, even if $\dot{M}_X$ does exist, arising if $x$ is anywhere non-differentiable. Instead, as with the transfer entropy, we must discuss \emph{pathwise} quantities, associated with complete path realisations, in order to associate active memory with individual behaviour. As such, we consider the \emph{accumulated} active memory utilisation on the interval $[t_0,t]$ to be $M_X^{[t_0,t]}$ which may be written
\begin{align}
M_X^{[t_0,t]}&=\int_{t_0}^{t}dt' \dot{M}_X(t')=\mathbb{E}\left[ \mathcal{M}^{[t_0,t]}_X[x_{[\tau,t]}]\right]
\end{align}
where $\mathcal{M}^{[t_0,t]}_X[x_{[\tau,t]}]$ is the \emph{pathwise active memory utilisation} on $[t_0,t]$, defined over complete paths as a logarithmic RN derivative between path measures
\begin{align}
 \mathcal{M}^{[t_0,t]}_X[x_{[\tau,t]}]&=\ln\frac{d\mathbb{P}_X[x_{(t_0,t]}|x_{[\tau,t_0]}]}{d\mathbb{P}^0_X[x_{(t_0,t]}|x_{t_0}]}\nonumber\\
 &\sim\lim_{\Delta t\to 0}\ln\prod_{i=0}^{n}\frac{p(x_{i+1}|x_{\{-k:i\}})}{p(x_{i+1}|x_{i})}
 \end{align}
 where $t_0=0$, $x_i=x_{i\Delta t}$, $n=(t/\Delta t)-1$ and $k=-\tau/\Delta t$. As with the transfer entropy, the pathwise active memory utilisation exists when the relevant RN derivative exists such that $\mathbb{P}_X$ and $\mathbb{P}_X^0$ are absolutely continuous with respect to each other. We denote the dynamics that emerge from the measure $\mathbb{P}_X^0$ the \emph{Markov marginal dynamics}. This in turn leads to the dual definition of the active memory utilisation rate
 \begin{align}
\dot{M}_X(t)&=\frac{d}{dt}\mathbb{E}\left[ \mathcal{M}^{[t_0,t]}_X[x_{[\tau,t]}]\right]\nonumber\\
&=\frac{d}{dt}\mathbb{E}\left[ \ln\frac{d\mathbb{P}_X[x_{(t_0,t]}|x_{[\tau,t_0]}]}{d\mathbb{P}^0_X[x_{(t_0,t]}|x_{t_0}]}\right].
\end{align}
Again, the alternative rate formulation
\begin{align}
\mathring{M}_X&=\lim_{t-t_0\to\infty}\frac{1}{t-t_0}\mathbb{E}\left[ \ln\frac{d\mathbb{P}_X[x_{(t_0,t]}|x_{[\tau,t_0]}]}{d\mathbb{P}^0_X[x_{(t_0,t]}|x_{t_0}]}\right]
\end{align}
behaves as $\mathring{M}_X=\dot{M}_X$ when the process is stationary.
If the process is both stationary and ergodic then this can be described through the expression
\begin{align}
\mathring{M}_X&=\lim_{t-t_0\to \infty}\frac{1}{t-t_0}\ln\frac{d\mathbb{P}_X[x_{(t_0,t]}|x_{[\tau,t_0]}]}{d\mathbb{P}^0_X[x_{(t_0,t]}|x_{t_0}]}
\end{align}
which is of use in empirical scenarios where an ensemble of realisations may not be available, but the process can be assumed to be stationary and ergodic.
\subsection{Instantaneous predictive capacity in continuous time}
\label{IXsec}
Here we describe the behaviour of the instantaneous predictive capacity asymptotically, illustrating this behaviour for distinct processes and relating it to the nature of the processes and their sampling paths. If the active information storage characterises the predictive capacity related to the prediction of some \emph{symbol} $x_{t+dt}$ that is stored in the history of $X$, the instantaneous predictive capacity characterises the residual part of this quantity once all predictive capacity related to the prediction of the transition \emph{event} $x_{t}\to x_{t+dt}$ has been identified (by $\dot{M}_X$). 
This instantaneous predictive capacity thus accounts for the predictive capacity of the symbol $x_{t+dt}$ `stored' in the current state $x_t$ and as such accounts for the reduction in uncertainty of the state $x_{t+dt}$ given instantaneous properties of the process such as its path regularity. Such a quantity has been defined in such a manner that it does not yield a rate and is thus not a dynamical quantity in the sense of $\dot{M}_X$ or $\dot{T}_{Y\to X}$. A corollary of this is that there exists no such analogous pathwise quantity $\mathcal{I}_X^{[t_0,t]}[x_{[\tau,t]}]$, like in the formulation of the active memory utilisation and transfer entropy, stemming from the non-existence of the proposed measure $\mathbb{P}_X^{\emptyset}$.
\par
Since Markov processes in continuous time possess no active memory utilisation, such processes have an instantaneous predictive capacity equal to their active information storage and thus are ideal for study here. Consequently, an example of instantaneous predictive capacity is the active information storage of the Ornstein Uhlenbeck process in Eq.~(\ref{AISOU}). We have seen that the active information storage, and thus instantaneous predictive capacity, diverges for such a process, but it is finite when considered as a prediction over some finite time $\Delta t$. As such we explore the idea that such a quantity can be characterised by the shape of the function with respect to $\Delta t$ in the vicinity of $\Delta t\to 0$.
\par
We do so by considering an asymptotic expansion of the instantaneous predictive capacity in the region $\Delta t\to 0$ of the prediction horizon and identify terms in different orders of $\Delta t$, similarly to the identification of distinct contributions in \cite{marzen_information_2014} (wherein the majority of contributions described behave much like $I_X$ since they cannot be formulated as RN derivatives and thus rates). As such we axiomatically identify these expected components of the instantaneous predictive capacity based on the following relationships
\begin{align}
I_X^I&=\lim_{\Delta t\to 0}I^{(\Delta t)}_X=I_X\nonumber\\
\dot{I}_X^R&=\lim_{\Delta t\to 0}\frac{ I^{(\Delta t)}_X-I_X^I}{\Delta t},
\end{align}
where analogously to $A_X^{(\Delta t)}$ in Eq.~(\ref{AISOU}), we define
\begin{align}
I^{(\Delta t)}_X&=\mathbb{E}\left[\ln\frac{p(x_{t+\Delta t}|x_t)}{p(x_{t+\Delta t})}\right].
\end{align}
We denote $\dot{I}_X^R$ the \emph{underlying instantaneous predictive capacity rate} and ${I}_X^I$ the \emph{non-dynamic instantaneous predictive capacity}. Assuming a common general asymptotic form \cite{norman_bleistein_asymptotic_1986} about $\Delta t=0$
\begin{align}
I^{(\Delta t)}_X&\sim \exp[-k\Delta t^{-\nu}]\sum_{i=0}^\infty\sum_{j=0}^{M(i)} c_{ij}(\ln{\Delta t})^{j}(\Delta t)^{r_i},\quad t\to 0^+,
\end{align}
with $k\geq 0$, $\nu>0$, $r_i\uparrow \infty$, we can then identify contributing components from the asymptotic expansion, which we observe to contribute for $k=0$, such that we have
\begin{align}
I_X^I&=\lim_{\Delta t\to 0}\sum_{i \forall r_i\leq 0}\sum_{j=0}^{M(i)} c_{ij}(\ln{\Delta t})^{j}(\Delta t)^{r_i},\nonumber\\
\dot{I}_X^R&=\lim_{\Delta t\to 0}\frac{1}{\Delta t}\sum_{i \forall r_i> 0}\sum_{j=0}^{M(i)} c_{ij}(\ln{\Delta t})^{j}(\Delta t)^{r_i}.
\end{align}
${I}_X^I$ has the same leading order behaviour as $I_X$ and characterises the contribution to the instantaneous predictive capacity not amenable to description as a rate, hence the characterisation as an instantaneous, i.e. non-infinitesimal, predictive quantity which has no dynamic analogue. $\dot{I}_X^R$ is thus the remaining leading order behaviour and characterises the continuous predictive influence of the process' history in the determination of future states as it dynamically evolves.
\par
These, quantities, however, are not guaranteed to be well defined. For instance ${I}_X^I$ converges iff $c_{ij}=0$ $\forall \; r_i<0, j>0$ and $\dot{I}_X^R$ converges iff $\min_{r_i>0} r_i=1$, and $c_{ij}=0$ $\forall \; r_i>0, j>0$. When such conditions are not met, the notion of an instantaneously held contribution and rate become undefined.
\par
It is instructive to examine such contributions for some simple processes. For the Ornstein-Uhlenbeck process detailed above we have $r_i=i$ and the following non-vanishing contributions for $i<2$
\begin{align}
c_{00}&=(1/2)\ln(1/(2\kappa))\nonumber\\
c_{01}&=-\frac{1}{2}\nonumber\\
c_{10}&=\frac{\kappa}{2},
\label{OUcoeff}
\end{align}
given by an expansion of Eq.~(\ref{AISOU}), revealing a divergent instantaneous contribution, but a well defined underlying rate since there are no $j>0$ $c_{1j}$ contributions.
\par
Considering, on the other hand, a process with different path regularity properties, for instance a master equation interpretation of the two species conversion process $A \underset{k_+}{\stackrel{k_-}{\rightleftharpoons}} B$ (with stationary solution $P_A=k_+/(k_-{+}k_+)$, $P_B=k_-/(k_-{+}k_+)$), shown in Appendix \ref{IXapp}, yields
\begin{align}
c_{00}&=-P_A\ln P_A-P_B\ln P_B\nonumber\\
c_{10}&=(k_-{+}k_+)^{-1}k_-k_+(\ln(k_-k_+)-2)\nonumber\\
c_{11}&=2(k_-{+}k_+)^{-1}k_-k_+.
\label{MEI}
\end{align}
Here we find a finite instantaneous contribution (equal to the Shannon information), yet an undefined rate.
\par
We can use these contributions to either assign a limiting instantaneous predictive capacity to each process, in these cases infinity for the Ornstein Uhlenbeck process and the Shannon entropy for the master equation process, or consider the instantaneous predictive capacity asymptotically and compare the contributions. For instance an Ornstein Uhlenbeck process with a smaller spring constant $\kappa$ has an asymptotically faster instantaneous contribution $I_X^I$, but a smaller underlying rate $I_X^R$.
\par
It is important to point out that these asymptotic terms demonstrate how $I_X$ is not due to \emph{memory} in any traditional sense, thus lending weight to the nomenclature that we have utilised. If, for example, we take the master equation process associated with Eqs.~(\ref{MEI}), the rates $k_+$ and $k_-$ might in reality be the Markov marginal rates of the true rates which may have some deep structure dependent on the past sequence of transitions such that one can predict a \emph{transition} with higher certainty based on this knowledge of the past. $I_X$ cannot detect this dependence and comprises the leading order contribution in $A_X$. Moreover, it thus follows $I_X$ or $A_X$ would therefore \emph{always} assign larger values to a Markov process without such deep structure merely on the basis of a larger instantaneous Shannon entropy, since the Shannon entropy is $\mathcal{O}(1)$ and the part of $A_X$ which can detect long range dependence, $\dot{M}_Xdt$, is $\mathcal{O}(dt)$.
\par
We point out that the convergence of these instantaneous predictive capacity contributions can be seen to be directly arising from the path regularity of each process. Processes that possess uncertainty in transitions along absolutely continuous sampling paths, naturally permit a rate of information `flow' from the history of the process. However, such processes are, by definition, defined in continuous space and possess vanishing uncertainty in the $\Delta t\to 0$ limit due to the absence of discontinuous transitions leading to an unmatched contribution in the form of a differential entropy of a delta function which diverges. On the other hand, processes with discrete states and sampling paths with a countable number of discontinuities, whilst similarly attaining vanishing uncertainty along their paths, possess a vanishing conditional Shannon entropy associated with transitions, leading to well a defined instantaneously held contribution. However, the path regularity which which affords such a well defined instantaneous contribution directly leads to $\ln(\Delta t)$ contributions arising exactly from the discontinuities, which render the notion of an underlying rate undefined.  In both cases the nature of the path regularity leads to logarithmic terms in $\Delta t$ in either the instantaneous, $I_X^I$, or rate, $\dot{I}_X^R$ contributions.

\section{Relation to other information theoretic concepts}
\subsection{Discrete time formalisms}
It is instructive to construct the quantities analogous to $I_X$ and $\dot{M}_X$ in discrete time and space in order to further illustrate the differences between them and what they quantify. In this case, at time $n$ with time origin $0$, we write
\begin{align}
A_X&=I_X+M_X\nonumber\\
{M}_X&=\mathbb{E}\left[\ln\frac{p(x_{n+1}|x_{\{0:n\}})}{p(x_{n+1}|x_n)}\right]\nonumber\\
I_X&=\mathbb{E}\left[\ln\frac{p(x_{n+1}|x_{n})}{p(x_{n+1})}\right],
\end{align}
where $M_X$ relates to $\dot{M}_X$ in the same way as the transfer entropy, $T_{Y\to X}$ relates to the transfer entropy rate $\dot{T}_{Y\to X}$. In this case $M_X$ is not a rate, but an $\mathcal{O}(1)$ quantity which may be thought of as the active memory utilisation associated with the time step $n\to n+1$. If we posit a process $X=x_i$ taking values $x_i\in\mathcal{X}=\{0,1,2,\ldots,N\}$, but at each time step only allow $x_i$ to transition to $x_{i+1}\in\{\{x_{i}+1,x_i,x_i-1\}\ \cap\ \mathcal{X}\}$, this constructs a process with a rudimentary path regularity property, dramatically restricting the space of complete paths $x_{\{0:n\}}$ that are realisable by the process. In this case, because time has been discretised, $A_X$ and $I_X$ are finite because the denominator, $p(x_{n+1})$, \emph{does} produce a probability measure, $\mathbb{P}_X^{\emptyset}$, over paths $x_{\{0:n\}}$; one where each time step is i.i.d. However, whilst $\mathbb{P}_X$ 
is absolutely continuous with respect to $\mathbb{P}_X^{\emptyset}$, 
 the two are not equivalent as $\mathbb{P}_X^{\emptyset}$ assigns probability to many paths that $\mathbb{P}_X$ does not, corresponding to paths that the process does not generate. Explicitly, $\mathbb{P}_X^{\emptyset}$ does not account for the property $x_{i+1}\in\{\{x_{i}+1,x_i,x_i-1\}\ \cap\ \mathcal{X}\}$, i.e. it will generate paths that can transition from any part of the phase space to any other in one time step. Consequently, in some steady state such that $p(x_i)>0 \ \forall\ x_i\in\mathcal{X}$, $A_X$ and $I_X$ get larger as $N$ gets larger without bound, since, as a rough approximation, it is measuring the relative size of $\mathcal{X}$ and $\{\{x_{i}+1,x_i,x_i-1\}\ \cap\ \mathcal{X}\}$. On the other hand, $M_X$ is constructed from measures that agree on which paths are possible and so does not have the unbounded dependence on the size of the state space. As such, one may loosely consider $M_X$ a property of storage associated with paths $x_{\{0:n\}}$,  independently of the nature of the ensemble in the wider phase space and its relation to any path regularity, whilst $I_X$ is a property of storage  characterising precisely this property which we may consider to be the relationship between the ensemble of paths $x_{\{0:n\}}$ and the ensemble of states $x_n$. In continuous time some path regularity is required for the process to exist and thus the latter component is not expressible as a rate.
\subsection{Information in continuous space and differential entropy}
\label{contdiff}
Here we provide an analogy between the issues that we have observed to arise markedly in continuous time and the well known issues surrounding the generalisation of Shannon entropy in continuous spaces and attempts to discuss it with differential entropy \cite{cover_elements_2005}. Given a continuous space, we may consider the information that can be stored as we increase our ability to resolve the space by partitioning into smaller and smaller regions. As we do this it becomes obvious that the information content of such a variable is, strictly, infinite, following directly from the arbitrary precision with which the variable in question can be specified. This is generally not a useful statement and as such, originally due to Shannon, the notion of differential entropy entered the field without any formal derivation, despite certain (grave) problems associated with it. Indeed it is not clear that such an object has any specific meaning in and of itself. On the other hand, \emph{relative} statements sidestep such issues, are well formed, finite, and are constructed with relative entropies frequently as Kullback-Leibler (KL) divergences. However, it is the exception and not the rule that events in a probability space of some variable are countable and have non-zero probability such that the variable possesses a simple Shannon entropy.
 As such it is not a particularly demanding claim that if one wants to construct robust, generalisable quantifications of random behaviour using information theory (for which we argue computational primitives of stochastic processes should be an example), one should \emph{always} be concerned with how different probability measures relate to each other naturally through KL divergences, which are mathematically underpinned by RN derivatives. For random processes, which are characterised by inter-related collections of random variables or random functions, if one wants to produce dynamic, finite quantities, the appropriate measures must concern the complete paths which the processes generate. An intuitive understanding for this might arise from an appreciation that if one does not compare the full probabilistic behaviour of paths over an interval, one will not be meaningfully creating a relative measure that accounts for the additional, and often infinite, precision (and thus information) that is available in the specification of complete random functions, or in the most simple case, in the \emph{timing} of events in continuous time.
\par
It is worth emphasising this point. Consider, for instance, a special case, where the random process, $X$, consists of a single impulse (with value $x_{I_t}=1$, where $I_t$ is the time of the impuse) in a window of length $\Delta t$ (where $x_t=0\;\forall\ t\neq I_t$). The phase space here is discrete, \{0,1\}: we cannot distinguish, and therefore cannot encode more than $\ln 2$ nats of information into the value of the symbol, but there is further variation that can be exploited due to its timing. The timing can occur at arbitrary precision and so we see that the entire process is functionally identical to a single, continuous, random variable $I_t\in[0,\Delta t]$ with a distribution of behaviour entirely captured by a one variable probability density, $p(I_t=t)$. It thus follows that, again, an infinite amount of information could be stored in such a process, indeed for \emph{any} duration $\Delta t$. This is an \emph{unavoidable} property of the process and, analogously, whilst one could construct a differential entropy to characterise this distribution, it would, necessarily, inherit all the problems associated with such a quantity in continuous space.
\par
If we examine the form of $I_X$ we can then understand why it is not equipped to balance this differential entropy in the sense of a KL divergence and thus return a convergent relative quantity. The numerator $p(x_{t+dt}|x_t)$ can be iteratively built up into a path probability density functionally identical to $p(I_t=t)$, which can be used to quantify the information content in both the variable $x_{t}$ \emph{and} its timing. However, information theoretic quantities based on the denominator, $p(x_t)$, are designed to quantify the information content in the single variable $x_t \in\{0,1\}$, i.e. the nature of the impulse. This form has no ability to relate such an event to the behaviour of the system at different times and so is agnostic to the timing of the impulse. This fundamental asymmetry is what causes $\mathcal{I}_X^{[t_0,t]}[x_{[\tau,t]}]$ and a rate of instantaneous predictive capacity to be ill defined and divergent respectively, a result that can loosely be interpreted as a quantification of the additional (infinite) information content the impulse process can leverage from its timing. 
\par
This does \emph{not} mean that quantities that return infinities, such as $\lim_{dt\to 0} A_X/dt$ and $\lim_{dt\to 0}  I_X/dt$ \footnote{We note the quantities $A_X$ and $I_X$ may be finite, and indeed meaningful, even if $A_X/dt$ or $I_X/dt$ are not.} are `incorrect' (we emphasise one, in theory, \emph{can} store infinite information in a continuous time process). Rather, in the context of complete paths in continuous time, they are probing answers to relatively unhelpful questions akin to asking the information content of a continuous variable. 
However, if we change the $p(x_t)$ terms that arise in the construction of these quantities to some other transition probability $p^*(x_{t+dt}|x_t)$, we are creating a relative measure that accounts for, or balances, this infinite precision in the timing, just as KL divergences on continuous spaces account for the infinite precision to which the symbols can be specified. This could correspond to questions such as `how much \emph{more} information can I store using one probability measure, or coding, over another'. This has a finite answer and is intimately related to our measure of memory utilisation. Again, each individual strategy confers \emph{infinite} information that can be encoded, but there is a finite relative measure.
\subsection{Excess entropy}
\label{excess}
A well known measure of information storage is the so-called excess entropy \cite{crutch03} which, for stationary processes, is a quantification of the shared or predictive information \cite{bialek_2001} between the semi-infinite past and future with expression as a mutual information
\begin{align}
E_X&=\lim_{k\to\infty}I[x_{\{n-k;n\}}; x_{\{n+1:n+k+1\}}]
\end{align}
in discrete time and
\begin{align}
E_X&=\lim_{r\to\infty}I[x_{[t-r,t]}; x_{(t,t+r]}].
\end{align}
in continuous time, where we naturally consider a time origin $\tau\to-\infty$. It should not be under-emphasised that the excess entropy possesses analogous properties in continuous time to the active information storage: in general (but not always) it cannot be expressed as a RN derivative between equivalent measures over paths. This can be seen by understanding that the excess entropy contains the active information storage. For instance,
it is known that, for a stationary process, in discrete time, the active information storage relates to the excess entropy as per \cite{liz10c,lizier_local_2012} 
\begin{align}
E_X&=\sum_{k=0}^\infty(A_X-A_X^{(k)})\nonumber\\
&=A_X+\sum_{k=1}^\infty(A_X-A_X^{(k)})
\end{align}
where
\begin{align}
A_X^{(k)}&=\mathbb{E}\left[\ln\frac{p(x_{n+1}|x_{\{n-k+1:n\}})}{p(x_{n+1})}\right],
\end{align}
with $A_X^{(k)}\leq A_X^{(k+1)}$ and $A_X^{(0)}=0$. 
In continuous time an analogous relation holds
\begin{align}
E_X&=A_X+\int_0^\infty ds\ \Delta \dot{A}_X^{(s)}
\end{align}
where
\begin{align}
\Delta \dot{A}_X^{(s)}&=\lim_{dt\to 0}\frac{1}{dt}({A}_X-{A}_X^{(s)})\geq 0\nonumber\\
{A}_X^{(s)}&=\mathbb{E}\left[\ln\frac{p(x_{t+dt}|x_{[t-s,t]})}{p(x_{t+dt})}\right].
\end{align}
Consequently, where $A_X$ is divergent, it follows that $E_X$ is too. 
Note, $\Delta \dot{A}_X^{(s)}$ is $\mathcal{O}(1)$, despite neither ${A}_X$ or ${A}_X^{(s)}$ individually leading to a well defined rate. 
This can be seen by noting that this expression is identical to
\begin{align}
E_X&=A_X+\int_0^\infty ds\ \Delta \dot{M}_X^{(s)}
\end{align}
where $\Delta \dot{M}_X^{(s)}=\Delta \dot{A}_X^{(s)}$, i.e.,
\begin{align}
\Delta \dot{M}_X^{(s)}&=\dot{M}_X-\dot{M}_X^{(s)}\nonumber\\
\dot{M}_X^{(s)}&=\lim_{dt\to 0}\frac{1}{dt}\mathbb{E}\left[\ln\frac{p(x_{t+dt}|x_{[t-s,t]})}{p(x_{t+dt}|x_t)}\right],
\end{align}
with both $\dot{M}_X$ and $\dot{M}_X^{(s)}$ expected to be convergent rates since they lead to integrated RN derivatives and where, typically $\lim_{s\to\infty}\dot{M}_X^{(s)}=\dot{M}_X$ such that $\int_0^\infty ds\ \Delta \dot{M}_X^{(s)}\in[0,\infty]$.  Consequently, whilst it may be common to observe $E_X=\infty$ and $A_X=\infty$ (e.g. for continuous processes such as the Ornstein-Uhlenbeck process and generalisations), their difference, $E_X-A_X$, may yet be a convergent $\mathcal{O}(1)$ quantity. And importantly, if it does not converge, it relates to the consideration of an infinite interval and behaviour in $\dot{M}_X^{(s)}$ which causes the appropriate integral on $[0,\infty]$ to be divergent, not the non-equivalence/non-existence of the measures under consideration.
\par
Somewhat reminicent of the discussion in Section \ref{contdiff}, again, we see a situation in which two information theoretic measures may be infinite, but possess  a \emph{relative} difference that is convergent. Explicitly, both $E_X$ and $A_X$ can be infinite, but possess a finite difference in the limit.  Indeed, since $A_X=I_X+\dot{M}_Xdt$ this  may, again in the limit, be written
\begin{align}
\lim_{dt\to 0}E_X-A_X&=\lim_{dt\to 0}E_X-I_X=\int_0^\infty ds\ \Delta \dot{M}_X^{(s)},
\label{elus}
\end{align}
implying that, more specifically, whenever $I_X$ diverges, $E_X$ does also such that a finite $E_X$ implies a finite (or vanishing) $I_X$.
Further, the quantity in Eq.~(\ref{elus}) exists in the literature as the so-called ``elusive'' information, \cite{james_anatomy_2011}, $\sigma_X=\lim_{s\to\infty}I[x_{(t,t+s]};x_{[t-s,t)}|x_t]$ since
\begin{align}
&\int_0^\infty ds\ \Delta \dot{M}_X^{(s)}\nonumber\\
&\qquad=\lim_{\substack{r\to\infty\\ dt\to 0}}\frac{1}{dt}\int_0^\infty ds\ \mathbb{E}\left[\ln\frac{p(x_{t+dt}|x_{[t-r,t]})}{p(x_{t+dt}|x_{[t-s,t]})}\right]\nonumber\\
&\qquad\simeq\lim_{\substack{n\to\infty\\ \Delta t\to 0\\r\to\infty}}\sum_{i=0}^{n}\mathbb{E}\left[\ln\frac{p(x_{t}|x_{[t-r,t]})}{p(x_{t}|x_{[t-i\Delta t,t]})}\right]\nonumber\\
&\qquad\simeq\lim_{\substack{n\to\infty\\ \Delta t\to 0\\r\to\infty}}\sum_{i=0}^{n}\mathbb{E}\left[\ln\frac{p(x_{t+(i+1)\Delta t}|x_{[t-r,t+i\Delta t]})}{p(x_{t+(i+1)\Delta t}|x_{[t,t+i\Delta t]})}\right]\nonumber\\
&\qquad=\lim_{r\to\infty}\mathbb{E}\left[\ln\frac{d\mathbb{P}_X[x_{(t,t+r]}|x_{[t-r,t]}]}{d\mathbb{P}_X[x_{(t,t+r]}|x_{t}]}\right],
\end{align}
having assumed stationarity in moving to line three. This quantity, being a logarithm of an RN derivative between equivalent measures, is $\mathcal{O}(1)$. This, in turn, clarifies and emphasises a different approach to information theoretic quantities in continuous time as opposed to that offered in, for example, \cite{marzen_information_2014}. In our approach one does not simply scale \emph{all} quantities one might consider by a time discretisation parameter, but instead identifies rates and integrated quantities, treating such quantities differently. The elusive information, like the expectation of the pathwise active memory utilisation and pathwise transfer entropy, is an integrated quantity associated with complete paths and as such is not meaningfully expressed as a rate with respect to a small time discretisation since it concerns behaviour that persists far beyond any such timescale. On the other hand, if the integral that characterises the elusive information does not converge, such that $\lim_{dt\to 0}E_X-I_X=\infty$, one can construct a rate, in the alternative sense, with respect to the entire process by defining, and considering in the limit $t\to\infty$,
\begin{align}
\mathring{\sigma}_X(t)=\lim_{r\to\infty}\frac{1}{t}I[x_{(t_0,t_0+t]};x_{[t_0-r,t_0)}|x_{t_0}],
\end{align}
or perhaps, when considering the explicit growth of the expected pathwise quantity that underlies $\sigma_X$, 
\begin{align}
\dot{\sigma}_X(t)&=\lim_{r\to\infty}\frac{d}{dt} I[x_{(t_0,t_0+t]};x_{[t_0-r,t_0)}|x_{t_0}]\nonumber\\
&=\dot{M}_X-\dot{M}_X^{(t)},
\end{align}
noting that these quantities are not equivalent. Nowhere, however, is the quantity $\lim_{dt\to 0}\sigma_X/dt$ implicated in the construction of such rates or expected to converge - for precisely the same reason it is not expected to for the quantities $T^{[t_0,t]}_{Y\to X}/dt$ or $M^{[t_0,t]}_{X}/dt$. Analogously, it follows, that $E_X$, $A_X$ and $I_X$ are natural $\mathcal{O}(1)$ quantities (though they may be infinite) and as such neither should we consider $\lim_{dt\to 0}E_X/dt$ etc.
\section{Components of information storage in jump and neural spiking processes}
With the preceding measures of instantaneous predictive capacity and active memory utilisation set out, we can describe such quantities in specific systems such that, when complemented by previous work on transfer entropy, a complete picture of information processing, as understood by the complementary description of memory and signalling, can be described. We acknowledge the parallel description of such processes, in the absence of extrinsic processes, in \cite{marzen_statistical_2016,marzen_informational_2017,marzen_structure_2017,marzen_time_2015-1}, which ultimately can be viewed as complementary in the sense of the connection between excess entropy and the quantities we consider as per Section \ref{excess}.
\subsection{Jump processes}
In previous work we described how to construct the relevant pathwise transfer entropy functional for jump processes for which neural spiking processes are a specific example \cite{spinney_transfer_2017}. Much of the resulting structure for active memory utilisation is analogous, but we reiterate the key points. We imagine, for simplicity, a discrete state space $x\in\mathcal{X}$. These are then stochastic processes characterised
by intermittent transitions between the states in $\mathcal{X}$ and
where the states are constant in-between these transitions. As such a path on the interval $[t_0,t]$, $x_{[t_0,t]}$, is characterised by the start and end times $t_0$ and $t$, its starting state $x_0$, and the times $t_i$ it transitions into the states $x_i$ such that we write $x_{[t_0,t]}\equiv \{t,\{t_i,x_i\}_0^{N_x}\}$ where $\{t_i,x_i\}_0^{N_x}\equiv\{t_0,x_0,t_1,x_1,\ldots,t_{N_x},x_{N_x}\}$ and where $N_x$ is the number of transitions in $x$ on the interval.
\par
Any given path realisation then possesses a probability density \cite{daley_introduction_2003}, constructed with the entire knowledge of the history of $x$, given some time origin $\tau$, at each point in time
\begin{align}
&p_X[x_{(t_0,t]}|x_{[\tau,t_0]}]=\nonumber\\
&\left(\prod_{i=1}^{N_x}W_X[x_{t_i}|x_{[\tau,t_i)}]\right)\exp{\left[-\int_{t_0}^t\lambda_X[x_{[\tau,t)}]\right]}
\end{align}
and a probability density constructed with knowledge only of the current value of $x$ at each point in time
\begin{align}
&p_X^0[x_{(t_0,t]}|x_{t_0}]\nonumber\\
&=\left(\prod_{i=1}^{N_x}W_X^0[x_{t_i}|x_{t_i}^-]\right)\exp{\left[-\int_{t_0}^t\lambda_X^0[x^{-}_{t}]\right]}
\end{align}
where $W$ are transition rates, $\lambda$ are escape rates and $x_{t}^- = \lim_{t'\nearrow t}x_{t'}$ such that
\begin{align}
W_X[x_{t_i}|x_{[\tau,t_i)}]&=\lim_{dt\to 0}\frac{1}{dt}p(x^-_{t_i}\to x_{t_i}\in[t_i,t_i+dt]|x_{[\tau,t_i)})\nonumber\\
W_X^0[x_{t_i}|x_{t_i}^-]&=\lim_{dt\to 0}\frac{1}{dt}p(x^-_{t_i}\to x_{t_i}\in[t_i,t_i+dt]|x_{t_i}^-)\nonumber\\
\lambda_X[x_{[\tau,t)}]&=\sum_{x'\neq x_t^-\in\mathcal{X}}W_X[x'|x_{[\tau,t)}]\nonumber\\
\lambda_X^0[x^{-}_{t}]&=\sum_{x'\neq x_t^-\in\mathcal{X}}W_X^0[x'|x_{t}^-].
\label{rates}
\end{align}
We note that such quantities may depend on the time $t_i$ or $t$ respectively should the process be non-stationary. The RN derivative that constitutes the pathwise active memory utilisation is then the ratio of these two quantities such that
\begin{align}
&\mathcal{M}^{[t_0,t]}_X[x_{[\tau,t]}]= \ln\frac{d\mathbb{P}_X[x_{(t_0,t]}|x_{[\tau,t_0]}]}{d\mathbb{P}^0_X[x_{(t_0,t]}|x_{t_0}]}\nonumber\\
&=\ln\frac{p_X[x_{(t_0,t]}|x_{[\tau,t_0]}]dt_1\ldots dt_{N_x}}{p_X^0[x_{(t_0,t]}|x_{t_0}]dt_1\ldots dt_{N_x}}\nonumber\\
&=\sum_{i=1}^{N_x}\ln\frac{W_X[x_{t_i}|x_{[\tau,t_i)}]}{W_X^0[x_{t_i}|x_{t_i}^-]}-\int_{t_0}^t(\lambda_X[x_{[\tau,t')}]-\lambda_X^0[x_{t'}])dt'.
\end{align}
As with the transfer entropy, there is a continuous integral component related to the waiting times between transitions and $N_x$ instantaneous contributions due to transitions between states. These instantaneous jumps are, in this instance, what stop a local rate (in the form of Eq.~(\ref{mlocrate})), from being well defined. Consequently we may decompose the total change of $\mathcal{M}^{[t_0,t]}_X[x_{[\tau,t]}]$ with time into these two components related to transitions $\Delta\mathcal{M}_X^{t}$ and waiting times $\dot{\mathcal{M}}_X^{nt}$, the latter of which \emph{does} permit a rate, such that
\begin{align}
\mathcal{M}^{[t_0,t]}_X[x_{[\tau,t]}]&=\sum_{i=1}^{N_x}\Delta\mathcal{M}_X^{t}(t_i)+\int_{t_0}^tdt'\;\dot{\mathcal{M}}_X^{nt}(t')\nonumber\\
\Delta\mathcal{M}^{t}_X(t_i)&=\ln{\frac{W_X[x_{t_i}|x_{[\tau,t_i)}]}{W_X^0[x_{t_i}|x_{t_i}^-]}}\nonumber\\
\dot{\mathcal{M}}^{nt}_X(t)&=\lambda_X^0[x^{-}_{t'}]-\lambda_X[x_{[\tau,t')}].
\end{align}
Importantly, since $\lambda_X^0[x^{-}_t]$ is just a marginalised average of $\lambda_X[x_{[\tau,t')}]$, when the expectation is taken we find
\begin{align}
\mathbb{E}[ \dot{\mathcal{M}}^{nt}_X(t)]&=\mathbb{E}\left[ \lambda_X[x_{[\tau,t)}]-\lambda_X^0[x^{-}_{t}]\right]\nonumber\\
&=0.
\label{meanwait0}
\end{align}
Consequently we may write
\begin{align}
\mathbb{E}\left[\mathcal{M}^{[t_0,t]}_X[x_{[\tau,t]}]\right]=\mathbb{E}\left[\sum_{i=1}^{N_x}\ln\frac{W_X[x_{t_i}|x_{[\tau,t_i)}]}{W_X^0[x_{t_i}|x_{t_i}^-]}\right],
\end{align}
and thus 
\begin{align}
\dot{M}_X(t)&=\frac{d}{dt}\mathbb{E}\left[\sum_{i=1}^{N_x}\ln\frac{W_X[x_{t_i}|x_{[\tau,t_i)}]}{W_X^0[x_{t_i}|x_{t_i}^-]}\right]\nonumber\\
&=\mathbb{E}\left[ (1-\delta_{x_{t},x_{t}^{-}})\ln\frac{W_X[x_{t}|x_{[\tau,t)}]}{W_X^0[x_{t}|x_{t}^-]}\right],
\end{align}
where $\delta_{x_{t},x_{t}^{-}}$ is the Kronecker delta. 
The alternative rate is given by
\begin{align}
\mathring{M}_X&=\lim_{t-t_0\to\infty}\frac{1}{t-t_0}\mathbb{E}\left[\sum_{i=1}^{N_x}\ln\frac{W_X[x_{t_i}|x_{[\tau,t_i)}]}{W_X^0[x_{t_i}|x_{t_i}^-]}\right]
\end{align}
equal to $\dot{M}_X$ when the process is stationary and we may write
\begin{align}
\mathring{M}_X&=\lim_{t-t_0\to\infty}\frac{1}{t-t_0}\sum_{i=1}^{N_x}\ln\frac{W_X[x_{t_i}|x_{[\tau,t_i)}]}{W_X^0[x_{t_i}|x_{t_i}^-]}.
\end{align}
when the process is ergodic.
\par
On the other hand, when considering the instantaneous predictive capacity, we emphasise, no analogous pathwise quantity $\mathcal{I}_X^{[t_0,t]}[x_{[\tau,t]}]$, and as discussed, no rate, exists for a direct comparison. However the mean rates $\dot{T}_{Y\to X}$ and $\dot{M}_X$ that emerge from this description sit alongside the asymptotic contributions to $I_X$. These contributions are obtained in Appendix \ref{IXapp}, and, for $X$ taking values in a set of discrete states $\mathcal{X}$, are given by 
\begin{align}
c_{00}&=-\sum_{x_t\in\mathcal{X}}P(x_t)\ln P(x_t)\nonumber\\
c_{10}&=\sum_{x^{-}_{t}\in\mathcal{X}}\sum_{x_{t}\neq x^{-}_{t} \in\mathcal{X}}P(x^{-}_{t})W_X^0[x_{t}|x^{-}_{t}]\nonumber\\
&\qquad\qquad\times\left[\ln\frac{W_X^0[x_{t}|x^{-}_{t}]P(x^{-}_{t})}{P(x_{t})}-1\right]\nonumber\\
c_{11}&=\sum_{x^{-}_{t}\in\mathcal{X}}\sum_{x_{t}\neq x^{-}_{t}\in\mathcal{X}}P(x^{-}_{t})W_X^0[x_{t}|x^{-}_{t}],
\end{align}
which is merely a generalisation of Eq.~(\ref{MEI}). As such we acknowledge the limiting value $I_X(t)$ is given by the Shannon entropy of the system at the time $t$.
\subsection{Neural spiking processes}
We consider an idealisation of neural processes whereby a realisation of the process consists entirely of indistinguishable, non overlapping, events (spikes) of duration $0$ seconds, such that any given path is characterised entirely by the timings of such events, $t_{i+1}>t_i$ etc., i.e. a point process. This can be constructed, in the sense of a stochastic process detailed above, in a number of ways, but here, rather than follow \cite{spinney_transfer_2017} where $X$ concerned the number of spikes that occurred since a time origin, we instead consider the limit of a two state system with the states, $0$ and $1$, corresponding to `not-spiked' and `spiked' respectively in the manner of burst noise or a telegraph process. Since there are only two states we have $ W_X[x_{t_i}|x_{[\tau,t_i)}]=\lambda_X[x_{[\tau,t_i)}]$ and $W_X^0[x_{t_i}|x_{t_i}^-]=\lambda_X^0[x_{t_i}^-]$. Finally, to achieve the reduction to a point process, we let $\lambda_X[x_{[\tau,t_i)}]_{x_{t_i}^-=1}=\lambda_X^0[x_{t_i}^{-}=1]$ such that there is no non-Markov dependence in the transition that characterises return to the unspiked state and then we consider the limit $\lambda_X^0[x_{t_i}^{-}=1]\to\infty$ such that the transition from the spiked state to the unspiked state is immediate. These two conditions ensure that there is no contribution to the active memory utilisation due to return to the unspiked state following spikes since
\begin{align}
\ln{\frac{\lambda_X[x_{[\tau,t_i)}]_{x_{t_i}^-=1}}{\lambda_X^0[x_{t_i}^{-}=1]}}=0
\end{align}
and that there is no contribution to the integral component from being in the spiked state since as $\lambda_X^0[x_{t_i}^{-}=1]\to\infty$, $x$ is in state $0$ with probability 1, such that
\begin{align}
&\int_{t_0}^t(\lambda_X[x_{[\tau,t')}]-\lambda_X^0[x^{-}_{t'}])dt'\nonumber\\
&\quad=\int_{t_0}^t(\lambda_X[x_{[\tau,t')}]_{x_{t'}^-=0}-\lambda_X^0[x_{t'}^-=0])dt'.
\end{align}
In this limit we may then characterise the process with a single transition/escape rate $\lambda_X[x_{[\tau,t)}]$, with expectation value $\lambda_X^0[x^{-}_{t}]=\lambda_X^0(t)$ which is understood to be the conditional spike rate with knowledge of the history of $x$ and the mean spike rate respectively. In addition, since the return transitions happen instantaneously we may now characterise the paths with simply the times of the spike events $x_{[t_0,t]}\equiv\{t,\{t_i\}_0^{N_x}\}$.
\par
Consequently, for neural spike processes we find
\begin{align}
\mathcal{M}^{[t_0,t]}_X[x_{[\tau,t]}]&=\sum_{i=1}^{N_x}\ln\frac{\lambda_X[x_{[\tau,t_i)}]}{\lambda_X^0(t_i)}\nonumber\\
&\quad-\int_{t_0}^t(\lambda_X[x_{[\tau,t')}]-\lambda_x^0(t'))dt',
\end{align}
which in turn possesses mean rates
\begin{align}
\mathring{M}_X&=\lim_{t-t_0\to\infty}\frac{1}{t-t_0}\mathbb{E}\left[\sum_{i=1}^{N_x}\ln\frac{\lambda_X[x_{[\tau,t_i)}]}{\lambda_X^0(t)}\right]\nonumber\\
\dot{M}_X(t)&=\mathbb{E}\left[ (1-\delta_{x_t,x_t^-})\ln\frac{\lambda_X[x_{[\tau,t)}]}{\lambda_X^0(t)}\right],
\end{align}
both equal for stationary processes and equal to
\begin{align}
\mathring{M}_X&=\lim_{t-t_0\to\infty}\frac{1}{t-t_0}\sum_{i=1}^{N_x}\ln\frac{\lambda_X[x_{[\tau,t_i)}]}{\lambda_X^0}
\end{align}
when the process is also ergodic.
\par
The instantaneous predictive capacity for spike processes, like with the active memory utilisation, is simply a special case of that for jump processes. Indeed, as formulated here, it is a special case of the two species conversion process in Section \ref{IXsec} with $A$ corresponding to the unspiked state and $B$ corresponding to the spiked state with rates $k^+=\lambda_X^0$ and $k^-=\mu$ in the limit of $\mu$ being taken to infinity such that we have
\begin{align}
c_{00}&=\lim_{p\to 1}-p\ln p -(1-p)\ln (1-p)\nonumber\\
&=0\nonumber\\
c_{10}&=\lim_{\mu\to\infty}(\lambda_X^0(t){+}\mu)^{-1}\lambda_X^0(t)\mu(\ln(\lambda_X^0(t)\mu)-2)\nonumber\\
&=\infty\nonumber\\
c_{11}&=\lim_{\mu\to\infty}2(\lambda_X^0(t){+}\mu)^{-1}\lambda_X^0(t)\mu\nonumber\\
&=2\lambda_X^0(t).
\label{IXspike}
\end{align}
As such we acknowledge the limiting value $I_X(t)=0$, noting that this can only emerge here as a consequence of assigning $0$ measure to the spiking phenomena on such time scales despite the fact that they occur almost surely in a sufficiently long time interval, and a divergent underlying rate, dominated by terms linear in the mean Markov intensity of the process. Indeed this limiting value, $I_X=0$, in conjuction with the corollary of Eq.~(\ref{elus}) that finite excess entropy, $E_X$, implies finite (or vanishing) instantaneous predictive capacity contextualises other results reporting convergent excess entropy in the case of point processes \cite{marzen_informational_2017,marzen_time_2015-1}.
\subsubsection{A note of caution with respect to empirical estimation techniques}
As with the estimation of the transfer entropy in such a setting \cite{spinney_transfer_2017}, we anticipate that the most efficient estimation of $\dot{M}_X$ in this context will emerge when utilising an estimator designed specifically to utilise inter-spike time intervals as relevant continuous variables. However, as with the transfer entropy a time binned approximation, whilst perhaps inefficient, will, in theory, be able to capture, in the limit, the behaviour of the above formalism. It is here, however, that spike train data can appear to be uniquely ambiguous when discretised in this fashion and for which we anticipate possible confusion if not performed carefully. For instance, if one attempts to discretise such that given a time interval $[0,t]$ with $N_x$ spikes based on a time resolution of $\Delta t$, such that one creates $N_x$ `spiked' bins and $(t/\Delta t)-N_x$ `not-spiked' bins one might appear to observe a convergent $A_X$ `rate' and vanishing $I_X$ `rate' by utilising $p(\text{spike})=N_x\Delta t/t$, calculating the relevant quantities and \emph{then} taking the limit. But this is incorrect as it is conflating $p(x_{t+dt})$ and $p(x_{t+dt}|x_t)$. Intuition as to the origin of the error in this hypothetical scheme can be gained by realising that there is no discretisation resolution, $\Delta t$, where a spike event takes up any more than one bin. Consequently \emph{all} finite probability associated with the spiked state is entirely an artefact of the discretisation procedure reflecting the correct probability $p(\text{spike})=0$. I.e. such a discretisation procedure would be conflating the `state'  of being spiked with associated vanishing probability, with the more appropriate characterisation of a spike as a \emph{transition} with associated probability density, with respect to a vanishing time interval, i.e. it would be interpreting $p(x_t)$ as a probability density, when it is not. 
As such, the quantity $N_x\Delta t/t$ reflects the probability of a spike any time \emph{within} an interval $\Delta t$, based on the history free statistics of the entire interval $[0,t]$, i.e. $\lambda_X^0\Delta t$ as per the definition of the transition (spike) rates in Eqs.~(\ref{rates}), which reflects the behaviour of $p(x_{t+dt}|x_t)$. Replacing $p(x_{t+dt})$ with $p(x_{t+dt}|x_t)$ in $A_X$  returns $\dot{M}_Xdt$ and thus the hypothetical naive calculations of a rate of $A_X$ would rather have been approximations of $\dot{M}_X$.
\subsection{Simple analytical examples}
\subsubsection{Stationary Poisson process with refractory period}
Perhaps the most simple, non-trivial, process, $X$ for which active memory utilisation is present in such a setting is a simple Poisson model of spiking with the introduction of a refractory period following any spike, of duration $\Delta_x$ seconds, during which the process cannot subsequently spike again. Defining $t_x$ to be the time of the most recent spike in $X$, and thus $t^x=t-t_x$ as the time \emph{since} the last spike in $X$, we may specify this through the conditional spike rate
\begin{align}
\lambda_X[x_{[\tau,t)}]&=\lambda_X(t^x)\nonumber\\
&=\begin{cases}
\mu & t^x \geq \Delta_x\\
0 & t^x<\Delta_x.
\end{cases}
\end{align}
Calculation of the relevant quantities can be achieved by exploiting the fact that the process is piece-wise Markovian between refractory periods and must appear Markovian and stationary when constructing the measure $\mathbb{P}_X^0$. Consequently, when we calculate $\lambda_X^0$ we can simply recognise that the characteristic time frame required to achieve a single spike following any previous spike is simply $\Delta_x +\mu^{-1}$ such that we have
\begin{align}
\lambda_X^0&=\frac{\mu}{1+\mu\Delta_x}.
\end{align}
Next we can then use these expressions to calculate the active memory utilisation rate contribution \emph{per spike}, due to that spike, varying only in its timing since the previous spike, given by
\begin{align}
&\lim_{t\to\infty}\int_{0}^{t}\exp\left[-\int_{t_x}^{t'}\lambda_X(t'')dt''\right]\lambda_X(t')\ln\frac{\lambda_X(t')}{\lambda_X^0}dt'\nonumber\\
&\quad=\int_{\Delta_x}^{\infty}\exp\left[-\mu(t-\Delta_x)\right]\mu\ln(1+\mu\Delta_x)dt\nonumber\\
&\quad=\ln(1+\mu\Delta_x).
\label{MPS}
\end{align}
There are then $\lambda_X^0$ of these spike events per unit time, each with the average contribution above. Further, since the contribution for non-spiking behaviour must vanish on average due to Eq.~(\ref{meanwait0}), it then follows that the active memory utilisation rate is
\begin{align}
\dot{M}_X&=\frac{\mu\ln(1+\mu\Delta_x)}{1+\mu\Delta_x}.
\end{align}
We see that $\mu$ serves, primarily, to control the number of spikes per unit time, thus scaling the number of prediction events and therefore the total scale of the active memory utilisation rate as a simple prefactor in addition to terms in $1+\mu\Delta_x$. 
Interestingly, however, this rate exhibits a maximum with respect to $\Delta_x$ for any given $\mu$ corresponding to $\Delta_x=\Delta^{\rm max}_x=(e-1)/\mu$ such that the largest active memory utilisation rate is given by $\dot{M}^{\rm max}_X=\mu/e$. Note that the form of these expressions, dependent on $e$, is not due to the choice of base of the logarithm used here. 
The existence of this maximum arises through the balance of two factors: increasing $\Delta_x$ increases the contribution \emph{per spike} through the logarithmic term as per Eq.~(\ref{MPS}), but also reduces the total number of expected spikes per unit time and thus total rate through the inverse $1+\mu\Delta_x$ term. We may also understand this phenomena through the distinguishability of the processes that correspond to $\mathbb{P}_X$ and $\mathbb{P}^0_X$. For values $\Delta_x\ll \Delta_x^{\rm max}$, the refractory period is not meaningfully impacting the dynamics such that both appear Poissonian. 
On the other hand,  values $\Delta_x \gg\Delta_x^{\rm max}$ also make the two processes, on aggregate, appear similar since, for the increasingly representative majority of the process (the refractory period), they behave as two processes with an arbitrarily low spike rate, despite the increase in mean contribution per spike. 
\par
Finally, as a spiking process, with a two value state space with vanishing Shannon entropy, the behaviour of $I_X$ is given by Eq.~(\ref{IXspike}).
\subsubsection{Non-stationary event driven spiking process}
\label{simple}
To illustrate the form of the active memory utilisation in a marginally more complicated  setting we use an elaborated simple toy model consisting of two spiking processes, $X$ and $Y$. 
We specify $Y$ to be a \emph{deterministic} spike train which spikes regularly with period $\Delta_y$. Noting that we have set the time origin $\tau=t_0$ for convenience, this means that the process always realises the specific path $y_{[t_0,t)}=y^{*}_{[t_0,t)}=\{\ldots,-\Delta_y,0,\Delta_y,2\Delta_y\ldots\}$, designed in this way to induce non-stationary behaviour in $X$. In this way, $Y$ could be considered to be some external stimuli occuring at regular time intervals, triggering separate trials in the sense of an event driven neural experiment. Next we specify that $X$ has a probability, $c$, of spiking within $\Delta_x$ seconds of the spike in $Y$ and that the spike occurs with uniform distribution on the interval $[t_y,t_y+\Delta_x]$ where $t_y\leq t$ is the time of the most recent spike in $Y$. Outside of this window $X$ cannot spike. We also insist on a refractory period of $\Delta_x$ seconds in the $X$ neuron during which it cannot spike immediately after spiking. Finally, we also specify that $2\Delta_x<\Delta_y$ such that  there can be no ambiguity in which spike in $Y$ is responsible for the possible spike in $X$ and only one spike in $X$ can result from any given spike in $Y$. A conditional spike rate that achieves this is given by
\begin{align}
&\lambda_{X|Y}[x_{[t_0,t)},y_{[t_0,t)}]\nonumber\\
&=\lambda_{X|Y}(t_x,t_y,t)\nonumber\\
&=\begin{cases}
\frac{c}{\Delta_x-c(t-t_y)} & t< t_y+\Delta_x\ \text{and}\ t\geq t_x+\Delta_x\\
0 & t\geq t_y+\Delta_x \ \text{or}\ t< t_x+\Delta_x,
\end{cases}
\label{Mratesimple}
\end{align}
where $t_x\leq t$ is the time of the last spike in $x$. This is easily observed since the probability of an absence of spikes on $[t_y,t]$, $t<t_y+\Delta_x$, is given by
\begin{align}
\exp{\left[-\int_{t_y}^{t}dt'\;\frac{c}{\Delta_x-c(t'-t_y)}\right]}&=1-\frac{c(t-t_y)}{\Delta_x},
\end{align}
such that the probability density of spiking at time $t$ is given by
\begin{align}
p(t)&=\frac{d}{dt}\frac{c(t-t_y)}{\Delta_x}=\frac{c}{\Delta_x},
\end{align}
i.e. a uniform distribution. Note that $\lambda_{X|Y}$ can be rewritten entirely in terms of the variables $t^x=t-t_x$ and $t^y=t-t_y$, the times \emph{since} the most recent spike in $X$ and $Y$, i.e. from variables encoded into the \emph{sequences} of the path histories of $X$ and $Y$ respectively, independently of the time, indicating that the behaviour encoded by $\lambda_{X|Y}$ is time homogeneous. Since $Y$ is deterministic, the form of the conditional spike rate in terms of $X$ is trivial. Noting the shorthand
\begin{align}
&\int dy_{[t_0,t)}p[x_{[t_0,t)},y_{[t_0,t)}]\nonumber\\
&\equiv p_{N_x,0}(t,\{t^x\}_0^{N_x})+\int_{t_0}^t dt^y_1\ p_{N_x,1}(t,\{t^x\}_0^{N_x},t^y_0)\nonumber\\
&+\int_{t_0}^t dt^y_1\int_{t_1^y}^t dt^y_2\ p_{N_x,2}(t,\{t^x\}_0^{N_x},\{t^y\}_0^2)\nonumber\\
&+\sum_{i=3}^{\infty}\int^t_{t_0}dt_1^y\ldots\int^t_{t_{i-1}^y}dt_i^y\ p_{N_x,i}(t,\{t_x\}_0^{N_x},\{t_y\}_0^i),
\label{shorth}
\end{align}
where $x_{[t_0,t)}\equiv \{t,\{t^x\}_0^{N_x}\}$, a path consisting of $N_x$ spikes, $p_{N_x,i}$ is the probability density of a joint spike sequence on $[t_0,t)$ with such $N_x$ spikes in $X$ and $i$ spikes in $Y$,  where $t_x^i$ and $t_y^i$ are the times of the $i$th spikes in $X$ and $Y$ respectively, and $t^x_0=t^y_0=t_0$, we may write
\begin{align}
&\lambda_X[x_{[t_0,t)}]\nonumber\\
&=\frac{1}{p[x_{[t_0,t)}]}\int dy_{[t_0,t)}\lambda_{X|Y}[x_{[t_0,t)},y_{[t_0,t)}]p[x_{[t_0,t)},y_{[t_0,t)}]\nonumber\\
&=\int dy_{[t_0,t)}\lambda_{X|Y}[x_{[t_0,t)},y_{[t_0,t)}]\frac{p[x_{[t_0,t)}|y_{[t_0,t)}]p[y_{[t_0,t)}]}{p[x_{[t_0,t)}]}\nonumber\\
&=\int dy_{[t_0,t)}\lambda_{X|Y}[x_{[t_0,t)},y_{[t_0,t)}]\nonumber\\
&\qquad\qquad\times\frac{p[x_{[t_0,t)}|y_{[t_0,t)}]}{p[x_{[t_0,t)}]}\delta(y_{[t_0,t)}-y^{*}_{[t_0,t)})\nonumber\\
&=\lambda_{X|Y}[x_{[t_0,t)},y^*_{[t_0,t)}]\frac{p[x_{[t_0,t)}|y^*_{[t_0,t)}]}{p[x_{[t_0,t)}]}\nonumber\\
&=\lambda_{X|Y}[x_{[t_0,t)},y^*_{[t_0,t)}],
\label{marg}
\end{align}
since $p[x_{[t_0,t)}]=\int dy_{[t_0,t)}p[x_{[t_0,t)}|y_{[t_0,t)}]\delta(y_{[t_0,t)}-y^*_{[t_0,t)})=p[x_{[t_0,t)}|y^*_{[t_0,t)}]$.
I.e. the joint process with a stationary dependence in $X$ on a deterministic $Y$ appears as a \emph{non-stationary} process in $X$ alone, such that
\begin{align}
\lambda_X[x_{[t_0,t)}]&=\lambda_X(t_x,t)\nonumber\\
&=\begin{cases}
\frac{c}{\Delta_x-c(t-n\Delta_y)} & t\in[n\Delta_y,n\Delta_y+\Delta_x]\\
& \text{and}\ t\geq t_x+\Delta_x\\
0 & t\notin[n\Delta_y,n\Delta_y+\Delta_x]\\
& \text{or}\ t< t_x+\Delta_x
\end{cases},
\end{align}
with $n\in\mathbb{Z}$. Note, in contrast to $\lambda_{X|Y}$, which can be written in terms of $t^x$ and $t^y$, $\lambda_X$ cannot be written solely in terms of $t^x$  and thus retains dependence on the time $t$, reflecting induced time inhomogeneity, leading to the observed non-stationary behaviour. This non-stationary behaviour is carried into $\lambda_X^0(t)$ which is given by considering the probability of a single spike in $[n\Delta_y,(n+1)\Delta_y]$, such that we need only consider $t_x\leq t-\Delta_x$, viz.
\begin{align}
\lambda_X^0(t)&=\frac{\Delta_x-c(t-n\Delta_y)}{\Delta_x}\lambda_X(t_x,t\in[n\Delta_y,n\Delta_y+\Delta_x])\nonumber\\
&\qquad+c\lambda_X(t_x,t\notin[n\Delta_y,n\Delta_y+\Delta_x])\nonumber\\
&=\begin{cases}
\frac{c}{\Delta_x}& t\in[n\Delta_y,n\Delta_y+\Delta_x]\\
0 & t\notin[n\Delta_y,n\Delta_y+\Delta_x],
\end{cases}
\label{l0}
\end{align}
such that the Markov marginal process cannot determine whether the process is in its refractory period or not. 
Since the process is non-stationary, it follows that $\dot{M}_X$ is not a constant here. Recognising, by construction, that we have a periodic process $Y$, an at most one to one mapping between spikes in $Y$ and $X$ and the property that outside of the refractory period, $X$ is piece-wise Markov, we may, without loss of generality, consider times $n\Delta_y\leq t< (n+1)\Delta_y$, treating $n\Delta_y$ as an effective time origin. Then, due to the refractory period, there is only a non-zero contribution for spike histories that have $0$ previous spikes in $[n\Delta_y,t]$, thus, again by construction, satisfying $t_x\leq t-\Delta_x$. Consequently, we may write
\begin{align}
\dot{M}_X(t)&=\mathbb{E}\left[(1-\delta_{x_t,x_{t}^-})\ln{\frac{\lambda_X[x_{[t_0,t)}]}{\lambda_X^0(t)}}\right]\nonumber\\\
&=\exp\left[-\int_{n\Delta_y}^tdt'\ \lambda_X(t_x\leq t-\Delta_x,t')\right]\nonumber\\
&\qquad\times\lambda_X(t_x\leq t-\Delta_x,t)\ln{\frac{\lambda_X(t_x\leq t-\Delta_x,t)}{\lambda_X^0(t)}}\nonumber\\
&=\begin{cases}\frac{c}{\Delta_x}\ln{\frac{\Delta_x}{\Delta_x-c(t-n\Delta_y)}}& t\in[n\Delta_y,n\Delta_y+\Delta_x]\\
0 & t\notin[n\Delta_y,n\Delta_y+\Delta_x],
\end{cases}
\end{align}
recovering the general case by once again letting $n\in\mathbb{Z}$. 
Further, we may consider $\mathring{M}_X$, which in this instance is simply 
\begin{align}
\mathring{M}_X&=\lim_{t\to \infty}\frac{1}{t}\int_0^t\dot{M}_X(t')dt'\nonumber\\
&=\frac{1}{\Delta_y}\int_{n\Delta_y}^{(n+1)\Delta_y}\dot{M}_X(t)dt\nonumber\\
&=\Delta_y^{-1}\left(c+(1-c)\ln(1-c)\right).
\end{align}
$\Delta_y$ serves simply to control the flow of predictable events and thus scale the total active memory utilisation rate, whilst $c$ controls how predictable each event is from the past of $X$, with the time dependence statistically identifiable in the conditioning in both the dynamics $\mathbb{P}_X$ and $\mathbb{P}_X^0$. In this case as $c$ increases, the more likely the refractory period is required to prevent a subsequent spike such that the processes characterised by $\mathbb{P}_X$ and $\mathbb{P}_X^0$ become more distinguishable leading to higher active memory utilisation. Again, as a spiking process, the behaviour of $I_X$ is given by Eq.~(\ref{IXspike}).
\par
We note that an intrinsic feature of this example, after $Y$ has been integrated out, has been its time inhomogeneity and subsequent non-stationary active memory utilisation. This has crucially been dependent on the notions that i) the stochastic behaviour of the process can be time inhomogeneous (implemented here through a \emph{deterministic} external process $Y$) and ii) this time inhomogeneity can be statistically detected. This amounts to an ability to determine conditional dependence upon the \emph{time} of evaluation such that, loosely, one can imagine that when conditioning on the sequence $x_{\mathcal{A}}$, one is always conditioning on both the history of $X$ \emph{and} the time, i.e. $\{x_{A},\mathcal{A}\}$ and that one can, in theory, draw multiple realisations of the process starting from the same time origin. This notion is explored and formalised in Appendix \ref{appA} along with a discussion of what one should expect if such time dependence existed, but the ability to either condition on the time or, equivalently, draw multiple realisations starting at the same time origin was unavailable. In short we find that if this is the case, one always over-estimates both the active memory utilisation and transfer entropy rates. This is demonstrated for the specific model considered here in Appendix section \ref{appA1} where we find that such an approach over-estimates the active memory utilisation rate by $(c/\Delta_y)\ln(\Delta_y/\Delta_x)$.
\subsection{Full information dynamics description of a neural spiking model}
Here we utilise a numerical model previously implemented in \cite{spinney_transfer_2017}, but also calculate the active memory utilisation alongside the transfer entropy, demonstrating their complementary nature, illustrating the behaviour of the pathwise quantities as a given pair of spike trains unfolds. In this model a spiking process, $Y$, follows a simple Poisson process characterised by a spike rate $\lambda_Y$ (note therefore, that $Y$ possesses an  active memory utilisation rate of $0$). Then the spiking process under consideration, $X$, spikes with a rate $\lambda_{X|Y}$ which depends upon the history of $Y$ uniquely through the time \emph{since} the last spike in $Y$, $t^y$,
\begin{figure*}[!htb]
\centering
\makebox[0pt]{\includegraphics[scale=0.5]{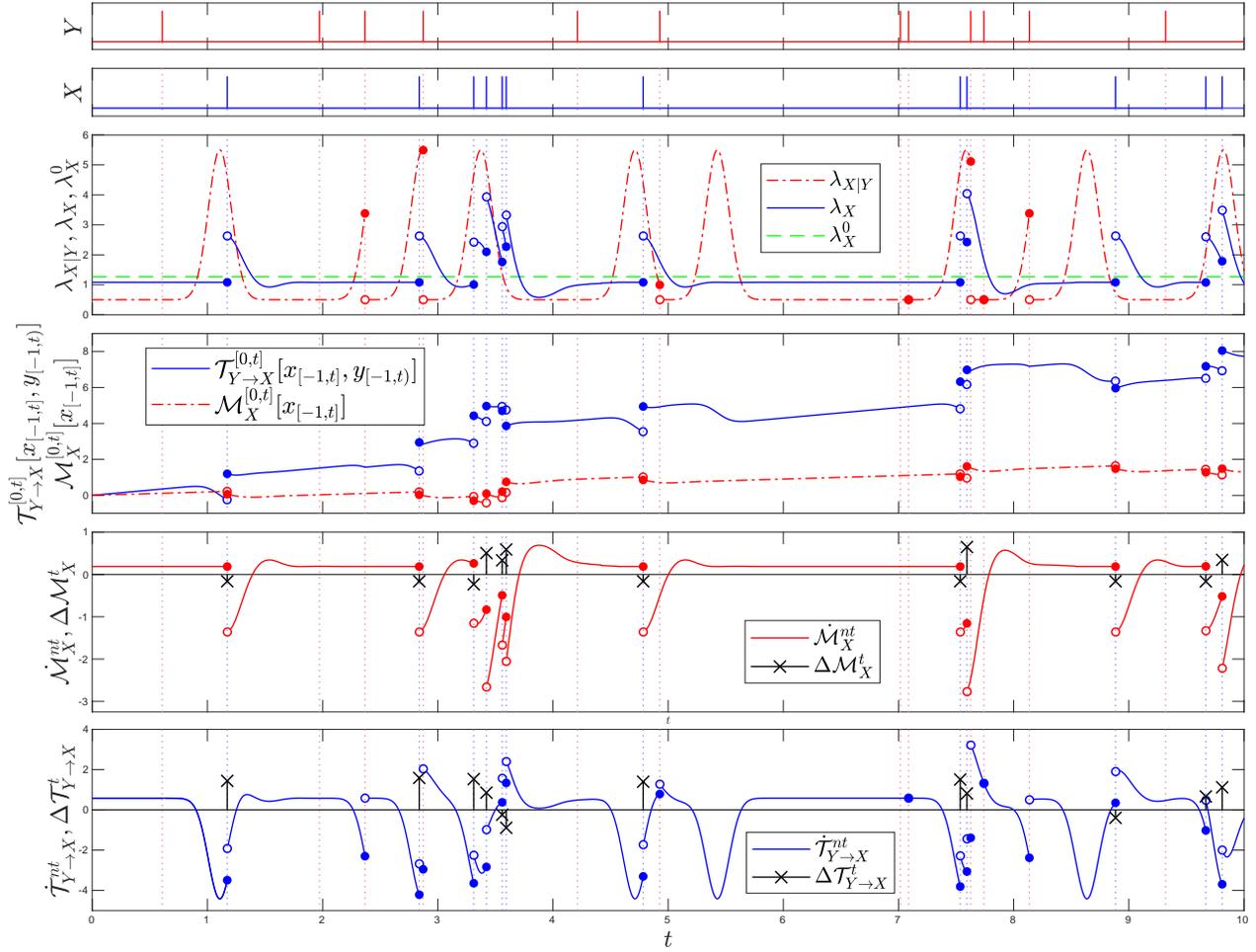}}
\caption{\label{fig2}Coupled spike trains generated using transition rates in Eq.~(\ref{sr}) using $\lambda_Y=1, \lambda_X^{\text{base}}=0.5, m=5, \sigma=0.1, t_{\text{cut}}=1$ along with generated and computed values of $\lambda_{X|Y}$, $\lambda_X$ \& $\lambda_X^0$, resulting pathwise transfer entropy ($\mathcal{T}^{[0,t]}_{Y\to X}[x_{[-1,t]},y_{[-1,t]}]$), pathwise active memory utilisation ($\mathcal{M}^{[0,t]}_{X}[x_{[-1,t]}]$) and local contributions ($\Delta \mathcal{T}_{Y\to X}^{t}$, $\dot{\mathcal{T}}_{{Y\to X}}^{nt}$, $\Delta \mathcal{M}_X^{t}$ and $\dot{\mathcal{M}}_X^{nt}$). We set $\tau=-1$, matching the maximum historical dependence in $\lambda_X$ and $\lambda_{X|Y}$, and a prior history of an absence of spikes in $Y$ and $X$ is assumed on the time interval $[-1,0)$. 
Sub figures $1$, $2$ and $6$ reprinted and sub figures $3$ and $4$ adapted with permission from [Richard E. Spinney, Mikhail Prokopenko and Joseph T. Lizier, Physical Review E, {\bf 95}, (3), 032319, (2017). \cite{spinney_transfer_2017}] Copyright 2017 by the
American Physical Society.}
\end{figure*}
\begin{align}
&\lambda_{X|Y}[x_{[\tau,t)},y_{[\tau,t)}]=\lambda_{X|Y}(t^y)\nonumber\\
&=\begin{cases}
\lambda_X^{\rm base} & t^y\notin (0,t_{\rm cut}],\\
\lambda_X^{\rm base}+m\exp\left[-\frac{1}{2\sigma^2}\left(t^y-\frac{t_{\rm cut}}{2}\right)^2\right] & t^y\notin (0,t_{\rm cut}],\\
-m\exp\left[-\frac{1}{2\sigma^2}\left(\frac{t_{\rm cut}}{2}\right)^2\right].&
\end{cases}
\label{sr}
\end{align}
The detailed dependence is illustrated in Fig.~(\ref{fig2}) where the behaviour of the pathwise active memory utilisation and transfer entropy are contrasted on the interval $[0,t]$, $t\in[0,10]$ assuming a time origin $\tau=-1$ with no spikes in the history of $X$ or $Y$ on $[-1,0]$ (not shown) and that the system is in its stationary state. Explicitly, given the spike rate detailed in Eq.~(\ref{sr}), we can identify (here in nats) the precise amount of information associated with memory utilisation, $\mathcal{M}^{[0,t]}_X[x_{[-1,t]}]$, and signalling $\mathcal{T}^{[0,t]}_{Y\to X}[x_{[-1,t]},y_{[-1,t)}]$ on an arbitrary interval $[0,t]$ for the specific spiking behaviour given in the first two subplots. 
The calculation of $\mathcal{M}^{[0,t]}_X[x_{[-1,t]}]$ relies upon the ability to compute $\lambda_X[x_{[-1,t)}]$ and $\lambda_X^0$ at all times whereas the calculation of $\mathcal{T}^{[0,t]}_{Y\to X}[x_{[-1,t]},y_{[-1,t)}]$ relies upon the ability to compute $\lambda_{X|Y}[x_{[-1,t)},y_{[-1,t)}]$ and $\lambda_X[x_{[-1,t)}]$ at all times. $\lambda_{X|Y}[x_{[-1,t)},y_{[-1,t)}]$ is specified by Eq.~(\ref{sr}), whilst $\lambda_X^0$, being a constant, since the process is stationary, is trivially determined by considering the mean number of spikes per unit time which is simply obtained by simulating the process and considering $\lambda_X^0=\lim_{t\to \infty} N_x/t$, where $N_x$ is the number of spikes in $[0,t]$, since the process is ergodic. Computation of $\lambda_X[x_{[0,t)}]$, however, is non-trivial, requiring a marginalisation integration over $\lambda_{X|Y}[x_{[-1,t)},y_{[-1,t)}]$ given the joint probabilistic behaviour of $\{x_{[-1,t)},y_{[-1,t)}\}$. An algorithm to compute this was reported in \cite{spinney_transfer_2017} and utilised here.

Information associated with other intervals $[t',t]$, $0<t'<t$ can be found by considering $\mathcal{M}^{[0,t]}_X[x_{[-1,t]}]-\mathcal{M}^{[0,t']}_X[x_{[-1,t']}]$ and $\mathcal{T}^{[0,t]}_{Y\to X}[x_{[-1,t]},y_{[-1,t)}]-\mathcal{T}^{[0,t']}_{Y\to X}[x_{[-1,t']},y_{[-1,t')}]$.
\par
Looking at how these quantities evolve in time, we see discontinuous contributions to both the pathwise transfer entropy and active memory utilisation when $X$ spikes with continuous contributions in between them. The contributions to transfer entropy and active memory utilisation are controlled by the relative sizes of $\lambda_{X|Y}$ and $\lambda_X$, and, $\lambda_{X}$ and $\lambda^0_X$ respectively. Positive discontinuous contributions in transfer entropy occur when $\lambda_{X|Y}>\lambda_X$ immediately preceding a spike in $X$ whilst positive continuous contributions occur when $\lambda_{X|Y}<\lambda_X$. Similarly, positive discontinuous contributions to active memory utilisation occur when $\lambda_{X}>\lambda^0_X$ immediately preceding a spike in $X$, whilst positive continuous contributions occur when $\lambda_{X}<\lambda^0_X$.
\par
One distinct difference in behaviour between  $\mathcal{M}^{[0,t]}_X[x_{[-1,t]}]$ and $\mathcal{T}^{[0,t]}_{Y\to X}[x_{[-1,t]},y_{[-1,t)}]$ can be observed in the evolution of the continuous contributions $\dot{\mathcal{M}}_X^{nt}$ and $\dot{\mathcal{T}}_{Y\to X}^{nt}$. $\dot{\mathcal{M}}_X^{nt}$ can only respond to changes in the history of $X$, whilst $\dot{\mathcal{T}}_{Y\to X}^{nt}$ can change in response to the history of both $X$ and $Y$. Consequently, for this particular system, we only see discontinuities in  $\dot{\mathcal{M}}_{ X}^{nt}$ when $X$ spikes, however we observe discontinuities in  $\dot{\mathcal{T}}_{Y\to X}^{nt}$ when either $X$ or $Y$ spikes since a spike in $X$ updates $\lambda_X$ whilst a spike in $Y$ updates $\lambda_{X|Y}$.
\par
 In the realisation specified above the majority of the predictive capacity which can be associated with the full paths is being derived from the additional reduction in uncertainty $Y$ provides through the transfer entropy, with a smaller residual predictive capacity being derived through the history of $X$ through the active memory utilisation.
\par
Finally, we briefly mention the behaviour of $I_X$, again given by Eq.~(\ref{IXspike}) with limiting value $I_X=0$. We can, however, for the particular numerical example in Fig.~(\ref{fig2}), state the numerical coefficient $c_{11}=2\lambda_X^0$ by reporting the mean Markov spike rate $\lambda_X^0\simeq 1.2697$.
\section{Information dynamics in generalised Ornstein-Uhlenbeck processes}
One of the more striking corollaries of the preceding formalism is that purely Markov processes like the Ornstein-Uhlenbeck process in Eq.~(\ref{OU}), whilst having non-zero, and indeed divergent, active information storage rates, have a vanishing memory utilisation rate since, by definition in their construction, they only have dependence on their most recent state. In many respects this is appealing as the memory utilisation rate then aligns very closely with the intuitive definition of a Markov process as being `memoryless'.
\par
However, if we couple multiple such Markov processes together, any individual process will no longer retain the Markov property due to the feedbacks between the processes. A simple example of such a model is that introduced in \cite{spinney_thermodynamics_2017} consisting of two linearly coupled Ornstein-Uhlenbeck processes with correlated noise viz.
\begin{figure*}[!htb]
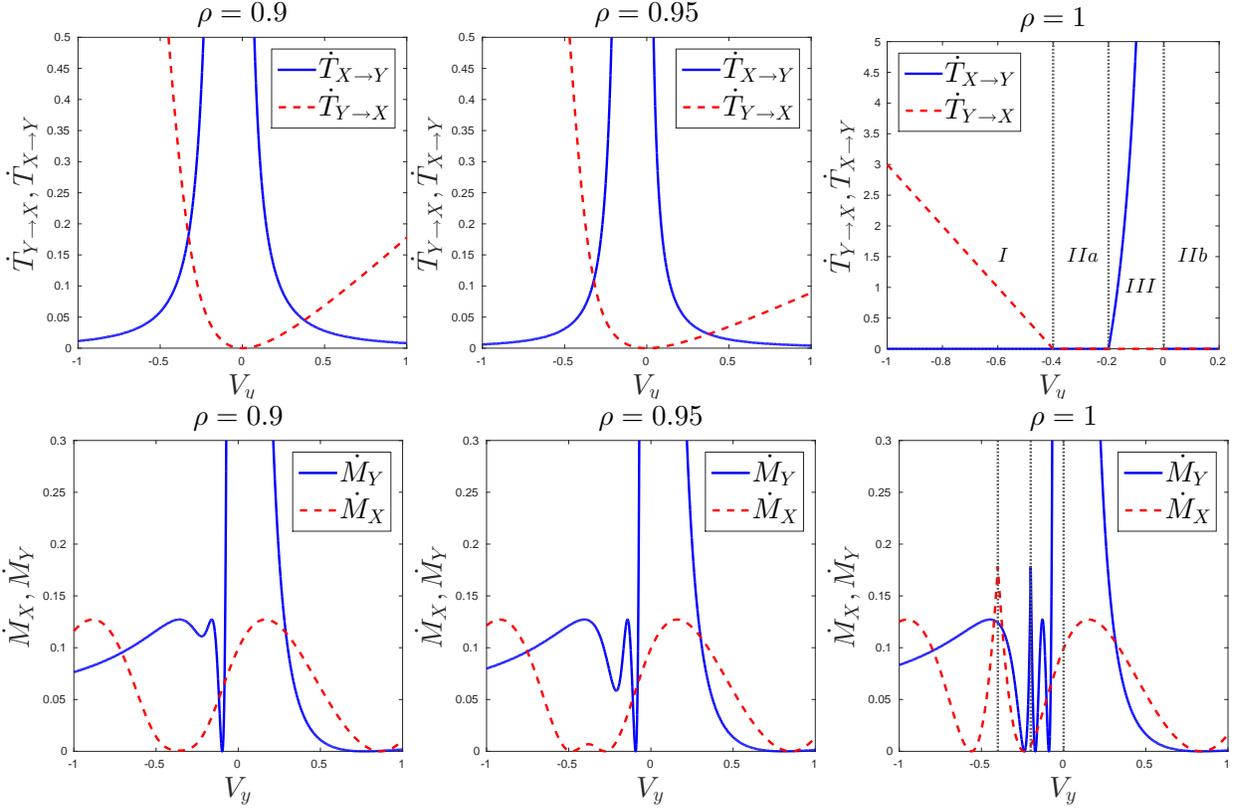

\includegraphics[width=0.3\textwidth]{OUAT_1}
\includegraphics[width=0.3\textwidth]{OUAT_2}
\includegraphics[width=0.3\textwidth]{OUAT_3}
\includegraphics[width=0.3\textwidth]{OUAT_4}
\includegraphics[width=0.3\textwidth]{OUAT_5}
\includegraphics[width=0.3\textwidth]{OUAT_6}
\caption{\label{fig3}Transfer entropy and active memory utilisation rates for the correlated coupled Ornstein-Uhlenbeck process given by Eqs.~(\ref{TEOU}) and (\ref{MTsum}) for varying noise correlation $\rho$ and noise strength in the $Y$ process $V_y$. We set $A=-5$, $B=5$, $C=1$, $D=-2$, $V_x=1$.}
\end{figure*}
\begin{align}
dx_t&=Ax_tdt+By_tdt+V_xdW^x_t\nonumber\\
dy_t&=Cx_tdt+Dy_tdt+V_ydW^y_t,
\label{sdes}
\end{align}
where $\mathbb{E}[ dW^x_tdW^y_{t'}]=\rho \delta(t-t')$ with $\rho\in[-1,1]$. The transfer entropy rate in the steady state of such a system is calculated in \cite{spinney_thermodynamics_2017} and given by
\begin{align}
&\dot{T}_{Y\to X}\nonumber\\
&=\frac{|D|}{2}\left(\sqrt{1+\frac{BV_y}{DV_x}\left(\frac{BV_y}{DV_x}-2\rho\right)}-\left(1+\rho\frac{BV_y}{|D|V_x}\right)\right).
\label{TEOU}
\end{align}
With the transfer entropy and all subsequent quantities, the symmetry of the process allows us to identify all analogous quantities associated with $Y$ ($\dot{T}_{X\to Y}$, $\dot{M}_Y$, $I_Y$) by making the substitutions $A\leftrightarrow D$, $B\leftrightarrow C$ and $V_x\leftrightarrow V_y$. To assess the full character of information processing in this system we wish to also consider $\dot{M}_x$ or, equivalently, $\dot{M}_x+\dot{T}_{Y\to X}$. The sum of these quantities is given by
\begin{align}
\dot{M}_X+\dot{T}_{Y\to X}&=\lim_{dt\to 0}\frac{1}{dt}\mathbb{E}\left[\ln\frac{p(x_{t+dt}|x_t,y_t)}{p(x_{t+dt}|x_t)}\right],
\end{align}
i.e. a Markov approximation to the transfer entropy given only the current values of the processes. 
\par
This reflects the property $\frac{d\mathbb{P}_{X|\{Y\}}}{d\mathbb{P}^0_{X}}=\frac{d\mathbb{P}_{X|\{Y\}}}{d\mathbb{P}_{X}}\frac{d\mathbb{P}_{X}}{d\mathbb{P}^0_{X}}$, since all the measures are equivalent, allowing us to write, in terms of pathwise quantities, 
\begin{align}
&\mathcal{M}_X^{[t_0,t]}[x_{[\tau,t]}]+\mathcal{T}_{Y\to X}^{[t_0,t]}[x_{[\tau,t]},y_{[\tau,t)}]\nonumber\\
&\qquad=\mathcal{M}_X^{[t_0,t]}[x_{[t_0,t]}]+\mathcal{T}_{Y\to X}^{[t_0,t]}[x_{[t_0,t]},y_{[t_0,t)}]\nonumber\\
&\qquad=\ln\frac {d\mathbb{P}_{X|\{Y\}}[x_{(t_0,t]}|x_{t_0},\{y_{[t_0,t)}\}]} {d\mathbb{P}_{X}^0[x_{(t_0,t]}|x_{t_0}]}.
\end{align}
Since both measures are Markovian and generate dynamics with the same sampling paths, with the same quadratic variation etc., we may thus leverage the Cameron-Martin-Girsanov theorem in recognising
\begin{align}
&\frac {d\mathbb{P}_{X|\{Y\}}[x_{(t_0,t]}|x_{t_0},\{y_{[t_0,t)}\}]} {d\mathbb{P}_{X}^0[x_{(t_0,t]}|x_{t_0}]}\nonumber\\
&=\exp\left[\frac{1}{2}\int_{t_0}^t f^2(x_{t'},y_{t'})dt'+\int_{t_0}^t f(x_{t'},y_{t'})dW^x_{t'}\right],
\label{gir}
\end{align}
such that
\begin{align}
&d\mathcal{M}_X^{[t_0,t]}[x_{[t_0,t]}]+d\mathcal{T}_{Y\to X}^{[t_0,t]}[x_{[t_0,t]},y_{[t_0,t)}]\nonumber\\
&\qquad\qquad=\frac{1}{2}f^2(x_t,y_t)dt+f(y_t,y_t)dW^x_t,
\end{align}
where
\begin{align}
f(x,y)=V_x^{-1}(Ax+By-\phi(x)),
\end{align}
and where $\phi(x)$ corresponds to the drift term in an effective dynamics which $\mathbb{P}_X^0$ describes, of the form
\begin{align}
dx_t&=\phi(x_t)dt+V_xdW_t^x.
\label{MMD}
\end{align}
Note, if we write $Z_t=\exp[-\mathcal{M}_X^{[t_0,t]}[x_{[t_0,t]}]-\mathcal{T}_{Y\to X}^{[t_0,t]}[x_{[t_0,t]},y_{[t_0,t)}]]$, it immediately follows that ${Z_t}$ is a Martingale, i.e.
\begin{align}
Z_t&=1-\int_{t_0}^tZ_{t'}f(x_{t'},y_{t'})dW^x_{t'},
\end{align}
implying $\mathbb{E}[ Z_t]=1$, not coincidentally mirroring the precise form of the so-called fluctuation theorems \cite{spinney_fluctuation_2013}, due to their analogous construction based on RN derivatives.
\par
The linear nature of the system dictates that the Markov marginal dynamics are also linear such that we have
\begin{align}
\phi(x)&=-\kappa_X^{\rm eff}x\nonumber\\
\kappa^{\rm eff}_X&=\frac{(A+D)(BC-AD)V_x^2}{D(A+D)V_x^2+B^2V_y^2-BV_x(CV_x+2\rho D V_y)},
\label{kxeff}
\end{align}
which can simply be read off the marginalised stationary distribution of the Fokker-Planck equation associated with the dynamics in Eq.~(\ref{sdes}) \cite{spinney_thermodynamics_2017} or found by marginalising the relevant short time transition probabilities/Greens functions and making appropriate manipulations using the stochastic calculus. 
From Eq.~(\ref{gir}) it thus follows that 
\begin{align}
&\dot{M}_X+\dot{T}_{Y\to X}\nonumber\\
&=\frac{1}{2}\mathbb{E}[ f^2(x,y)]\nonumber\\
&=\left(4(A+D)V_x^2\right)^{-1}\nonumber\\
&\times\left(B^2V_y^2+D(A+D)V_x^2-BV_x(2D\rho V_y+CV_x)\right)^{-1}\nonumber\\
&\times B^2\left(-B^2V_y^4+2B(D-A)\rho V_y^3V_x\right.\nonumber\\
&\left.\qquad\quad-((A+D)^2-2BC-4AD\rho^2)V_x^2V_y^2\right.\nonumber\\
&\left.\qquad\quad-2C(D-A)\rho V_yV_x^3-C^2 V_x^4\right),
\label{MTsum}
\end{align}
with the active memory utilisation rate being the difference between Eqs.~(\ref{MTsum}) and (\ref{TEOU}). The comparison of such terms yields rich structure even in such a simple system. Some of this structure is shown in Fig.~(\ref{fig3}) where the approach to complete correlation in the noise terms is shown for different noise strengths in $Y$. We note that both the transfer entropy and active memory utilisation rate, $\dot{T}_{X\to Y}$ and $\dot{M}_Y$, diverge whenever $V_y\to 0$ since in this limit $Y$ becomes a deterministic process such that the conditions required on RN derivatives for their existence are not met. In particular we point out the behaviour when $\rho=1$ where we observe different regimes in the character of information processing, marked in the top right sub figure. In regime I $\dot{T}_{Y\to X}>0$ whilst $\dot{T}_{X\to Y}=0$ and in regime III the opposite behaviour is observed, $\dot{T}_{X\to Y}>0$ whilst $\dot{T}_{Y\to X}=0$. In the remaining regimes IIa and IIb all the transfer entropy rates are zero. The transitions between these regimes are marked with fine dashed lines. Interestingly, at the transitions separating regimes I and IIa and IIa and III we see a sharp discontinuous, but not divergent, peak in the accompanying active memory utilisation rates. The critical value of the noise strength in $Y$ that separates regimes I and IIa is given by $V_y^{\rm crit}=DV_x/B$, whilst the critical value of the noise strength in $Y$ that separates regimes IIa and III is given by $V_y^{\rm crit}=CV_x/A$.
\par
Since the Markov mariginal process is characterised by Eq.~(\ref{MMD}), specified with Eq.~(\ref{kxeff}), ie. a simple Ornstein-Uhlenbeck process as in Eq.~(\ref{OU}), the contributions to the instantaneous predictive capacity, $I_X$, and components $I_X^I$ and $\dot{I}_X^R$, are equal to those in Eq.~(\ref{OUcoeff}), but with $\kappa=\kappa^{\rm eff}_X$ from Eq.~(\ref{kxeff}).
\section{Discussion and Conclusions}
In this paper we have  extended the approach elaborated in \cite{spinney_transfer_2017} for treating transfer entropy in continuous time to the broader framework of information dynamics. In doing so we have decomposed the active information storage into two distinct, positive, quantities called the active memory utilisation and instantaneous predictive capacity. The former is complementary to the transfer entropy and inherits much of the behaviour of transfer entropy in continuous time: there is a central cumulative quantity, the pathwise active memory utilisation, associated with finite time intervals which possesses a mean rate at single instances in time. Individual behaviour, or events, are characterised by the pathwise active memory utilisation, rather than a local rate, since the pathwise active memory utilisation may be discontinuous. The latter quantity, the instantaneous predictive capacity, retains the predictive capacity of the process which does not assign finite contributions to individual path realisations, accounting for fundamental properties of the process such as the continuity properties of its sampling paths. Further, we have offered an asymptotic formalism for discussing this contribution, highlighting key differences in its structure for different processes. Since it accounts for intrinsic properties such as path continuity, which may lead to infinite predictive capacities, but has been derived in the context of separately maximising all finite, pathwise, positive contributions, it is the minimal measure capable of offsetting similar effects in other measures of information processing, such as the excess entropy. Doing so reveals the maximum constituent component constructed from a cumulative pathwise quantity in the excess entropy is the so-called elusive information, which we have clarified should not be treated as a rate, but as an $\mathcal{O}(1)$ quantity independently of the time basis.
\par
We have then constructed such a formalism in the context of jump and neural spiking processes, complementing our previous work \cite{spinney_transfer_2017}. Using this we have demonstrated how to assess the complete information processing occurring in such a context comprising both memory and signalling. This has been illustrated in synthetic models of neural spiking demonstrating the qualitative behaviour one should expect. Further, we have shown that the concepts offered here are well defined in other processes including coupled Ornstein-Uhlenbeck processes where we report novel and interesting fine structure in the interplay between memory and signalling.
\par
As with the transfer entropy, this work offers great promise particularly within the field of computational neuroscience where such a formalism lends itself to efficiently quantifying information processing in such settings.
We finish by highlighting two particular consequences of our work in a neuroscience setting. First, that as per estimation of the transfer entropy for neural spike trains \cite{spinney_transfer_2017}, we anticipate that the most efficient estimation of $\dot{M}_X$ for such processes will emerge when utilising an estimator utilising inter-spike time intervals (which completely describe the process) as relevant continuous variables. Second, that where active information storage is measured on discrete-time samples of underlying continuous-time processes (as is the case in neural imaging measurements), the active memory utilisation is the only component that will approach a limiting value as the discrete time step approaches zero, and so may be the most appropriate quantity for investigation in such experiments.
\acknowledgements
The authors thank Mikhail Prokopenko for useful discussions that contributed to this work.
J.L. was supported through the Australian Research Council DECRA Grant No. DE160100630, and a Faculty of Engineering and IT Early Career Researcher and Newly Appointed Staff Development Scheme grant.
\bibliographystyle{apsrev4-1}
\appendix
\section{Selected derivations of results}
\label{appA0}
\subsection{Active information storage of the Ornstein-Uhlenbeck process}
\label{appA01}
The process described by Eq.~(\ref{OU}) permits a well known solution to its transition probability by consideration of the Green's function of the corresponding Fokker-Planck equation, due to the method of characteristics  \cite{risken_fokker-planck_1996}, given by
\begin{align}
p^{(\Delta t)}_{OU}(x_{t+\Delta t}|x_t)&=\sqrt{\frac{\kappa}{\pi\sigma^2(1-e^{-2\kappa\Delta t})}}\nonumber\\
&\quad\times\exp{\left[-\frac{\kappa(x_{t+\Delta t}-x_te^{-\kappa \Delta t})^2}{\sigma^2(1-e^{-2\kappa\Delta t})}\right]},
\end{align}
consistent with the stationary solution
\begin{align}
p_{OU}(x_{t})&=\sqrt{\frac{\kappa}{\pi\sigma^2}}\exp{\left[-\frac{\kappa x_{t}^2}{\sigma^2}\right]} \quad\forall\ t.
\end{align}
The active information storage is then simply given by the integral
\begin{align}
A_X^{(\Delta t)}&=\int_{-\infty}^{\infty}dy\int_{-\infty}^{\infty}dx\ p^{(\Delta t)}_{OU}(y|x)p_{OU}(x)\ln{\frac{p^{(\Delta t)}_{OU}(y|x)}{p_{OU}(y)}},
\end{align}
leading to the result in Eq.~(\ref{AISOU}).
\subsection{Instantaneous predictive capacity of discrete state master equations}
\label{IXapp}
Given the master equation on the discrete state space $\mathcal{X}$,
\begin{align}
\dot{P}(x_i)&=\sum_{x_j\neq x_i\in\mathcal{X}}W[x_i|x_j]P(x_j)-W[x_j|x_i]P(x_i),
\end{align}
we can calculate the instantaneous predictive capacity over a short time $\Delta t$ by considering up to $1$ transition such that
\begin{align}
I_X^{(\Delta t)}&=\sum_{x_i\in\mathcal{X}}\sum_{x_j\neq x_i \in\mathcal{X}}P(x_i)\left[W[x_j|x_i]\Delta t\ln\frac{W[x_j|x_i]\Delta t}{P(x_j)}\right.\nonumber\\
&\left.\qquad\quad+(1-\lambda[x_i]\Delta t)\ln\frac{(1-\lambda[x_i]\Delta t)}{P(x_i)}\right]+\mathcal{O}(\Delta t^2),
\end{align}
revealing the contributions
\begin{align}
c_{00}&=-\sum_{x_i\in\mathcal{X}}P(x_i)\ln P(x_i)\nonumber\\
c_{10}&=\sum_{x_i\in\mathcal{X}}P(x_i)\Bigg[\lambda[x_i]\left(\ln P(x_i)-1\right)\nonumber\\
&\qquad\qquad\qquad+\sum_{x_j\neq x_i \in\mathcal{X}}W[x_j|x_i]\ln\frac{W[x_j|x_i]}{P(x_j)}\Bigg]\nonumber\\
&=\sum_{x_i\in\mathcal{X}}\sum_{x_j\neq x_i \in\mathcal{X}}P(x_i)W[x_j|x_i]\left[\ln\frac{W[x_j|x_i]P(x_i)}{P(x_j)}-1\right]\nonumber\\
c_{11}&=\sum_{x_i\in\mathcal{X}}\sum_{x_j\neq x_i\in\mathcal{X}}P(x_i)W[x_j|x_i].
\end{align}
The instantaneous predictive capacity for the two species conversion process,  $A \underset{k_+}{\stackrel{k_-}{\rightleftharpoons}} B$, utilised in Section \ref{IXsec} can be derived from the dynamics underlying the master equation
\begin{align}
\dot{P}_A&=k_-P_B-k_+P_A\nonumber\\
\dot{P}_B&=k_+P_A-k_-P_B,
\end{align}
i.e. $\mathcal{X}=\{A,B\}$, $W[A|B]=\lambda[B]=k_-$, $W[B|A]=\lambda[A]=k_+$ and with stationary solution $P_A=k_-/(k_-+k_+)$, $P_B=1-P_A=k_+/(k_-+k_+)$ yielding the coefficients in Eq.~(\ref{MEI}).
\section{Conditional and non-stationary variants}
\label{appA}
Based on the formulation presented in the main text it is trivial to generalise, both the transfer entropy and active memory utilisation, to the conditional case. In discrete time the conditional forms, conditioned upon some third variable $Z$ are given by
\begin{align}
T_{Y\to X|Z}&=\mathbb{E}\left[\ln\frac{p(x_{n+1}|x_{\{0:n\}},y_{\{0:n\}},z_{\{0:n\}})}{p(x_{n+1}|x_{\{0:n\}},z_{\{0:n\}})}\right]\nonumber\\
M_{X|Z}&=\mathbb{E}\left[\ln\frac{p(x_{n+1}|x_{\{0:n\}},z_{\{0:n\}})}{p(x_{n+1}|x_{n},z_{\{0:n\}})}\right]
\end{align}
which is straightforward to generalise to the continuous time case based on the generalisation of the pathwise transfer entropy and pathwise active memory utilisation to the \emph{conditional pathwise transfer entropy} and \emph{conditional pathwise active memory utilisation}
\begin{align}
&\mathcal{T}^{[t_0,t]}_{Y\to X|Z}[x_{[\tau,t]},y_{[\tau,t)},z_{[\tau,t)}]=\nonumber\\
&\quad\ln\frac{d\mathbb{P}_{X|\{Y,Z\}}[x_{(t_0,t]}|x_{[\tau,t_0]},\{y_{[\tau,t)},z_{[\tau,t)}\}]}{d\mathbb{P}_{X|\{Z\}}[x_{(t_0,t]}|x_{[\tau,t_0]},\{z_{[\tau,t)}\}]}\nonumber\\
&\mathcal{M}^{[t_0,t]}_{X|Z}[x_{[\tau,t]},z_{[\tau,t)}]=\nonumber\\
&\quad\ln\frac{d\mathbb{P}_{X|\{Z\}}[x_{(t_0,t]}|x_{[\tau,t_0]},\{z_{[\tau,t)}\}]}{d\mathbb{P}^0_{X|\{Z\}}[x_{(t_0,t]}|x_{[\tau,t_0]},\{z_{[\tau,t)}\}]}
\end{align}
where, analogously to the above, we may consider
\begin{align}
&\frac{d\mathbb{P}_{X|\{Y,Z\}}[x_{(t_0,t]}|x_{[\tau,t_0]},\{y_{[\tau,t)},z_{[\tau,t)}\}]}{d\mathbb{P}_{X|\{Z\}}[x_{(t_0,t]}|x_{[\tau,t_0]},\{z_{[\tau,t)}\}]}
\sim\nonumber\\
&\qquad\quad\lim_{n\to\infty}\prod_{i=0}^n\frac{p(x_{i+1}|x_{\{-k:i\}},y_{\{-k:i\}},z_{\{-k:i\}})}{p(x_{i+1}|x_{\{-k:i\}},z_{\{-k:i\}})}\nonumber\\
&\frac{d\mathbb{P}_{X|\{Z\}}[x_{(t_0,t]}|x_{[\tau,t_0]},\{z_{[\tau,t)}\}]}{d\mathbb{P}^0_{X|\{Z\}}[x_{(t_0,t]}|x_{[\tau,t_0]},\{z_{[\tau,t)}\}]}
\sim\nonumber\\
&\qquad\quad\lim_{n\to\infty}\prod_{i=0}^n\frac{p(x_{i+1}|x_{\{-k:i\}},z_{\{-k:i\}})}{p(x_{i+1}|x_{i},z_{\{-k:i\}})},
\end{align}
where, again, $t_0=0$, $x_i\equiv x_{i\Delta t}$ and $\Delta t=t/(n+1)=-\tau/k$. Again we note the general construction of path probabilities with the $\{\}$ notation to mean $\mathbb{P}_{X|\{A\}}[x_{(t_0,t]}|x_{[\tau,t_0]},\{A_{[\tau,t)}\}]\sim \prod_{i=0}^np(x_{i+1}|x_{\{-k:i\}},A_{\{-k:i\}})$ where $A$ is some arbitrary extrinsic variable or variables in the form of a coincident time series.
\par
This is, perhaps, not so illuminating in general, however it allows us to be precise when we discuss non-stationary transfer entropy and active memory utilisation rates. To this end we make it clear that whenever such quantities are calculated, the time at which any transition probability is evaluated is also known such that one can construct them as conditional variants, conditioned on a third, deterministic, `variable', $\mathfrak{T}_t$, taking values $t'_t$ equal to the time it is indexed by, i.e. $t'_t=t$.  Thus, by conditioning on $\mathfrak{T}$ each relevant probability measure identifies any time dependence. As such, we explicitly take the following statements to be synonymous
\begin{align}
\mathcal{T}^{[t_0,t]}_{Y\to X}[x_{[\tau,t]},y_{[\tau,t)}]&\equiv\mathcal{T}^{[t_0,t]}_{Y\to X|\mathfrak{T}}[x_{[\tau,t]},y_{[\tau,t)},t'_{[\tau,t)}]\nonumber\\
\mathcal{M}^{[t_0,t]}_{X}[x_{[\tau,t]}]&\equiv\mathcal{M}^{[t_0,t]}_{X|\mathfrak{T}}[x_{[\tau,t]},t'_{[\tau,t)}]
\end{align}
such that it is only for brevity that $\mathfrak{T}$ is omitted in their formulation. We note that since $\mathfrak{T}$ merely represents the time index this alters some of the properties associated with the conditional probabilities used to construct the RN derivatives, namely that $t'_{\mathcal{A}}=\mathcal{A}$ and conditioning on $[\tau,t]$ is identical to conditioning on $t$. 
Accordingly we may write the RN derivatives underlying the quantities as
\begin{align}
&\exp\left[\mathcal{T}^{[t_0,t]}_{Y\to X|\mathfrak{T}}[x_{[\tau,t]},y_{[\tau,t)},t'_{[\tau,t)}]\right]\nonumber\\
&\qquad=\frac{d\mathbb{P}_{X|\{Y,\mathfrak{T}\}}[x_{(t_0,t]}|x_{[\tau,t_0]},\{y_{[\tau,t)},[\tau,t)\}]}{d\mathbb{P}_{X|\mathfrak{T}}[x_{(t_0,t]}|x_{[\tau,t_0]},\{[\tau,t)\}]}\nonumber\\
&\qquad\sim\lim_{n\to\infty}\prod_{i=0}^n\frac{p(x_{i+1}|x_{\{-k:i\}},y_{\{-k:i\}},i\Delta t)}{p(x_{i+1}|x_{\{-k:i\}},i\Delta t)}\nonumber\\
&\exp\left[\mathcal{M}^{[t_0,t]}_{ X|\mathfrak{T}}[x_{[\tau,t]},t'_{[\tau,t)}]\right]\nonumber\\
&\qquad=\frac{d\mathbb{P}_{X|\{\mathfrak{T}\}}[x_{(t_0,t]}|x_{[\tau,t_0]},\{[\tau,t)\}]}{d\mathbb{P}^0_{X|\{\mathfrak{T}\}}[x_{(t_0,t]}|x_{[\tau,t_0]},\{[\tau,t)\}]}\nonumber\\
&\qquad\sim\lim_{n\to\infty}\prod_{i=0}^n\frac{p(x_{i+1}|x_{\{-k:i\}},i\Delta t)}{p(x_{i+1}|x_{i},i\Delta t)}.
\end{align}
It is common, however, to assume stationarity in a process when these quantities are computed empirically. This amounts to using the alternative measures $\mathbb{P}^{\rm st}_{X|\{Y\}}$, $\mathbb{P}^{\rm st}_{X}$ and $\mathbb{P}^{0,{\rm st}}_{X}$, constructed by averaging the probabilistic behaviour experienced at all observed times, assuming equal a priori probabilities with respect to any instance in time. This means that the estimated quantities converge to different `stationary' quantities denoted $\mathcal{T}^{{\rm st}, [t_0,t]}_{Y\to X}[x_{[\tau,t]},y_{[\tau,t]}]$ and $\mathcal{M}^{{\rm st}, [t_0,t]}_{X}[x_{[\tau,t]}]$, and, $\dot{T}^{\rm st}_{Y\to X}$ and $\dot{M}^{\rm st}_X$ defined through
\begin{align}
&\mathcal{T}^{{\rm st}, [t_0,t]}_{Y\to X}[x_{[\tau,t]},y_{[\tau,t]}]=\ln\frac{d\mathbb{P}^{\rm st}_{X|\{Y\}}[x_{(t_0,t]}|x_{[\tau,t_0]},\{y_{[\tau,t)}\}]}{d\mathbb{P}^{\rm st}_{X}[x_{(t_0,t]}|x_{[\tau,t_0]}]}\nonumber\\
&\sim\lim_{n\to\infty}\ln\left[\prod_{i=0}^n\frac{p^{\rm st}(x_{i+1}|x_{\{-k:i\}},y_{\{-k:i\}})}{p^{\rm st}(x_{i+1}|x_{\{-k:i\}})}\right]\nonumber\\
&=\lim_{n\to\infty}\ln\left[\prod_{i=0}^n\lim_{n\to\infty}\frac{\sum_{j=-k}^{n}p(x_{i+1}|x_{\{-k:i\}},y_{\{-k:i\}},j\Delta t)}{\sum_{j=-k}^{n}p(x_{i+1}|x_{\{-k:i\}},j\Delta t)}\right]\nonumber\\
&\mathcal{M}^{{\rm st}, [t_0,t]}_{X}[x_{[\tau,t]}]=\ln\frac{d\mathbb{P}^{\rm st}_{X}[x_{(t_0,t]}|x_{[\tau,t_0]}]}{d\mathbb{P}^{0,{\rm st}}_{X}[x_{(t_0,t]}|x_{[\tau,t_0]}]}\nonumber\\
&\sim\lim_{n\to\infty}\ln\left[\prod_{i=0}^n\frac{p^{{\rm st}}(x_{i+1}|x_{\{-k:i\}})}{p^{{\rm st}}(x_{i+1}|x_i)}\right]\nonumber\\
&=\lim_{n\to\infty}\ln\left[\prod_{i=0}^n\lim_{n\to\infty}\frac{\sum_{j=-k}^{n}p(x_{i+1}|x_{\{-k:i\}},j\Delta t)}{\sum_{j=-k}^{n}p(x_{i+1}|x_i,j\Delta t)}\right].
\end{align}
I.e. we understand that the empirically computed probabilities, assuming stationarity, will approximate
\begin{align}
&p^{\rm st}(x_{i+1}|x_{\{-k:i\}})\nonumber\\
&\quad=\lim_{n\to \infty}\frac{1}{n+k+1}\sum_{j=-k}^np(x_{i+1}|x_{\{-k:i\}},t=j\Delta t).
\end{align}
Clearly, however, if only one or a limited number of samples are available, this approximation cannot be expected to accurate, in the general case, however long the samples, unless, for example, the underlying time variation in $p$ is periodic or, if controlled by some hidden variable, that variable evolves ergodically.  
As such, only when a process is stationary do we explicitly have $\mathcal{T}^{{\rm st}, [t_0,t]}_{Y\to X}=\mathcal{T}^{[t_0,t]}_{Y\to X}$ and $\dot{T}^{\rm st}_{Y\to X}=\dot{T}_{Y\to X}$, and, $\mathcal{M}^{{\rm st}, [t_0,t]}_{X}=\mathcal{M}^{[t_0,t]}_{X}$ and $\dot{M}^{\rm st}_{X}=\dot{M}_{X}$. Moreover, we have formulated the difference between these quantities in terms of a marginalisation over an implied process, $\mathfrak{T}$. Consequently, treating this process like any other, we can identify the difference between the `stationary' and non-stationary quantities as
\begin{align}
\dot{T}^{\rm st}_{Y\to X}-\mathring{T}_{Y\to X}&=\dot{T}_{\mathfrak{T}\to X}-\dot{T}_{\mathfrak{T}\to X|Y}
\end{align}
and
\begin{align}
\dot{M}^{\rm st}_X-\mathring{M}_{X}&=\dot{T}^{(0)}_{\mathfrak{T}\to X}-\dot{T}_{\mathfrak{T}\to X}
\label{MgtM}
\end{align}
where, analogously,
\begin{widetext}
\begin{align}
\dot{T}_{\mathfrak{T}\to X}&=\lim_{t\to\infty}\frac{1}{t-t_0}\int_{t_0}^{t}\lim_{dt\to 0}\frac{1}{dt}\mathbb{E}\left[\ln\frac{p(x_{t'+dt}|x_{[\tau,t')},t')}{p^{\rm st}(x_{t'+dt}|x_{[\tau,t')})}\right] dt'\nonumber\\
&=\lim_{t\to\infty}\frac{1}{t-t_0}\int_{t_0}^{t}\lim_{dt\to 0}\frac{1}{dt}\mathbb{E}\left[\ln\frac{p(x_{t'+dt}|x_{[\tau,t')},t')}{\lim_{t\to\infty}(t-t_0)^{-1}\int_{t_0}^t p(x_{t'+dt}|x_{[\tau,t')},t'')dt''}\right] dt'\nonumber\\
\dot{T}^{(0)}_{\mathfrak{T}\to X}&=\lim_{t\to\infty}\frac{1}{t-t_0}\int_{t_0}^{t}\lim_{dt\to 0}\frac{1}{dt}\mathbb{E}\left[\ln\frac{p(x_{t'+dt}|x_{t'},t')}{p^{\rm st}(x_{t'+dt}|x_{t'})}\right] dt'\nonumber\\
&=\lim_{t\to\infty}\frac{1}{t-t_0}\int_{t_0}^{t}\lim_{dt\to 0}\frac{1}{dt}\mathbb{E}\left[\ln\frac{p(x_{t'+dt}|x_{t'},t')}{\lim_{t\to\infty}(t-t_0)^{-1}\int_{t_0}^t p(x_{t'+dt}|x_{t'},t'')dt''}\right] dt'.
\end{align}
\end{widetext}
In particular, if $\dot{T}_{\mathfrak{T}\to X}=0$, such that an infinite history length in $X$ allows us predict equally well with or without a time index, we risk over-estimation of the active memory utilisation since $\dot{M}^{\rm st}_X-\mathring{M}_{X}=\dot{T}^{(0)}_{\mathfrak{T}\to X}\geq 0$.  
\par
It is important to note that it may be challenging, at least empirically, to identify such a distinction between stationary and non-stationary formulations, as it requires us to be able to probe the statistics of the process at a given time $t$. The ability to achieve this requires i) the ability to hypothetically draw multiple samples or realisations from the generating process starting at the same time origin ii) implicitly allow access to the time of evaluation when considering any transition probability. Indeed this may be deemed completely impossible in practice leaving such a question fundamentally ambiguous. 
\subsection{Stationary active memory utilsation calculation for the model utilised in Section \ref{simple}}
\label{appA1}
We can illustrate the above distinctions and possible ambiguity by calculating $\dot{M}_X^{\rm st}$ for the process described in Section \ref{simple} and discuss when and how the distinction between the two calculations would be the same, different or unknowable. In this system, non-stationary behaviour is introduced by means of a deterministic variable $Y$ upon which the behaviour in $X$ depends. The system can be identified as non-stationary by appealing to an ensemble of realisations at any given time $t$. We can, however, imagine that the time indexing of such an ensemble is either not known or not knowable, equivalent to the assumption that the system is stationary if the spike rates were to be constructed empirially from data.
\par
To calculate $\dot{M}_X^{\rm st}$, therefore, requires calculation of two analogous spike rates $\lambda_X^{\rm st}$ and $\lambda_X^{{\rm st},0}$ where only the sequences of the relevant path histories are known, and not the time at which they are being evaluated. Calculation of $\lambda_X^{{\rm st},0}$ is straight forward and amounts to the aggregated mean spike rate in $X$, which can either just be asserted by recognising that there are, on average, $c\cdot(T/\Delta_y)$ spikes in an interval $T$ in the $T\to\infty$ limit or by writing
\begin{align}
\lambda_X^{{\rm st},0}&=\lim_{t-t_0\to\infty}\frac{1}{t-t_0}\int_{t_0}^t\lambda_X^0(t')dt\nonumber\\
&=\frac{1}{\Delta_y}\int_{n\Delta_y}^{(n+1)\Delta_y}\lambda_X^0(t')dt'\nonumber\\
&=\frac{c}{\Delta_y}.
\end{align} 
On the other hand, $\lambda_X^{\rm st}$ depends on how much path history is available to condition upon. We will take the limit of a time origin $\tau=t_0 \to-\infty$, both for simplicity and because such a condition  will dominate in the case $t\to\infty$ when considering long time, steady state, behaviour. We find such behaviour in several steps. First, noting again the shorthand of Eq.~(\ref{shorth}), we construct
\begin{align}
&\lambda^{\rm st}_X[x_{[t_0,t)}]\nonumber\\
&=\frac{1}{p^{\rm st}[x_{[t_0,t)}]}\int dy_{[t_0,t)}\lambda^{\rm st}_{X|Y}[x_{[t_0,t)]},y_{[t_0,t)}]p^{\rm st}[x_{[t_0,t)},y_{[t_0,t)}]\nonumber\\
&=\int dy_{[t_0,t)}\lambda_{X|Y}[x_{[t_0,t)},y_{[t_0,t)}]\frac{p^{\rm st}[x_{[t_0,t)}|y_{[t_0,t)}]p^{\rm st}[y_{[t_0,t)}]}{p^{\rm st}[x_{[t_0,t)}]}
\end{align}
where we have recognised $\lambda_{X|Y}^{\rm st}[x_{[t_0,t)},y_{[t_0,t)}]=\lambda_{X|Y}[x_{[t_0,t)},y_{[t_0,t)}]$ due to the ability to write $\lambda_{X|Y}[x_{[t_0,t)},y_{[t_0,t)}]$ independently of $t$ as per Section \ref{simple}. Next we consider the form of $p^{\rm st}[y_{[t_0,t)}]$. Recall $Y$ always realises $y^*_{[t_0,t)}=\{\ldots,-\Delta_y,0,\Delta_y,2\Delta_y,\ldots\}$ such that $p[y_{[t_0,t)}]=\delta(y_{[t_0,t)}-y^*_{[t_0,t)})$. $p^{\rm st}[y_{[t_0,t)}]$ is then the probability of the sequence of $y^*_{[t_0,t)}$ disassociated with its timing such that we do not know if we are considering the probability of the \emph{sequence} $y^*_{[t_0,t)}$ at time $t$ or any other time. Consequently, $p^{\rm st}[y_{[t_0,t)}]$ assigns  probability to every path $y_{[t_0,t)}$ that consists of spikes at precise  intervals of $\Delta_y$, but is shifted by an arbitrary phase factor $\phi\in[0,\Delta_y)$, i.e. $y_{[t_0,t)}=y^*_{[t_0+\phi,t+\phi)}$, according to the  distribution $p_{\Phi}(\phi)$ (which is necessarily flat given equal a priori probability at all times, though left general for completeness and later discussion). I.e. all variation in $y_{[t_0,t)}$ is through the single parameter $\phi$. Consequently we have $p^{\rm st}[x_{[t_0,t)},y_{[t_0,t)}]=p^{\rm st}[x_{[t_0,t)}|\phi]p_{\Phi}(\phi)$ and thus

\begin{align}
\lambda^{\rm st}_X[x_{[t_0,t)}]&=\int_0^{\Delta_y} d\phi\ \lambda_{X|Y}[x_{[t_0,t)},\phi]\frac{p^{\rm st}[x_{[t_0,t)}|\phi]p_{\Phi}(\phi)}{p^{\rm st}[x_{[t_0,t)}]}.
\end{align}
At this point we make the critical claim
\begin{align}
\lim_{t_0\to -\infty}p^{\rm st}[x_{[t_0,t)}|\phi]p_{\Phi}(\phi)=\lim_{t_0\to -\infty}p^{\rm st}[x_{[t_0,t)}]\delta(\phi)
\label{limlam}
\end{align}
which is to say, given a long enough past sequence $x_{[t_0,t)}$ there is only a single compatible value of $\phi$. Since the process is actually generating $y_{[t_0,t)}^*$ then this value must be $\phi=0$. This can be seen by posing the (inverse) question: given an arbitrarily long sequence $x_{[t_0,t)}$ (along with knowledge of its time indexing) does there exist a way of uniquely determining $\phi$? The answer is yes with it being obtained by searching the past sequence of $X$ for the minimum interspike interval. In the infinite limit this is the smallest \emph{possible} interspike interval. This occurs when the first of such spikes coincides with the very end of a possible spiking window following a spike in $Y$, with timing $t=n\Delta_y +\Delta_x+\phi$, and the subsequent spike occuring at the very beginning of the next possible spiking window following a spike in $Y$, with timing $t=(n+1)\Delta_y+\phi$ (with $n\in\mathbb{Z}$). Consequently
\begin{align}
&\lim_{t_0\to-\infty}\lambda^{\rm st}_X[x_{[t_0,t)}]\nonumber\\
&\qquad=\lim_{t_0\to-\infty}\int_0^{\Delta_y}d\phi\ \lambda_{X|Y}[x_{[t_0,t]},\phi] \frac{p^{\rm st}[x_{[t_0,t]}]}{p^{\rm st}[x_{[t_0,t]}]}  \delta(\phi)\nonumber\\
&\qquad=\lim_{t_0\to-\infty}\lambda_{X|Y}[x_{[t_0,t]},\phi=0]\nonumber\\
&\qquad=\lambda_{X|Y}[x_{[t_0,t]},y^*_{[\tau,t)}]\nonumber\\
&\qquad=\lambda_{X|Y}(t_x,t_y)\nonumber\\
&\qquad=\lambda_{X}(t_x,t),
\label{lamst}
\end{align}
i.e. all the predictive capability of $\lambda_{X|Y}$ (and thus $\lambda_X$) can be gleaned from the history of $X$ even without knowledge of the current time. 
\par
We then use these two stationary transition rates to construct the average associated with the calculation of $\dot{M}_X^{\rm st}$. However, since $\lambda_X^{{\rm st},0}$ is just a constant and $\lim_{t_0\to-\infty}\lambda^{\rm st}_X[x_{[t_0,t)}]=\lambda_X(t_x,t)$, this allows us to greatly simplify the implied average over all infinitely long paths $x_{[t_0,t)}$ by replacing $\lambda^{\rm st}_X[x_{[t_0,t)}]$ with $\lambda_X(t_x,t)$ and integrating over a single period $[n\Delta_y,(n+1)\Delta_y]$. As such we find
\begin{align}
&\dot{M}^{\rm st}_X\nonumber\\
&=\lim_{t_0\to-\infty}\mathbb{E}\left[(1-\delta_{x_t,x_{t}^-})\ln{\frac{\lambda^{\rm st}_X[x_{[t_0,t)}]}{\lambda_X^{{\rm st},0}}}\right]\nonumber\\
&=\lim_{t_0\to-\infty}\int dx_{[t_0,t)}p^{\rm st}[x_{[t_0,t)}]\lambda^{\rm st}_X[x_{[t_0,t)]}]\ln{\frac{\lambda^{\rm st}_X[x_{[t_0,t)}]}{\lambda_X^{{\rm st},0}}}\nonumber\\
&=\frac{1}{\Delta_y}\int_{n\Delta_y}^{(n+1)\Delta_y}\mathbb{E}\left[(1-\delta_{x_t,x_{t}^-})\ln{\frac{\lambda_X(t_x,t)}{\lambda_X^{{\rm st},0}}}\right] dt\nonumber\\
&=\frac{1}{\Delta_y}\int_{n\Delta_y}^{(n+1)\Delta_y}\exp\left[-\int_{n\Delta_y}^t\lambda_X(t_x\leq t-\Delta_x,t')dt'\right]\nonumber\\
&\quad\times\lambda_X(t_x\leq t-\Delta_x,t)\ln{\frac{\Delta_y\lambda_X(t_x\leq t-\Delta_x,t)}{c}}dt\nonumber\\
&=\frac{1}{\Delta_y}\int_{n\Delta_y}^{n\Delta_y+\Delta_x}\frac{\Delta_x-c(t-n\Delta_y)}{\Delta_x}\nonumber\\
&\quad\times\lambda_X(t_x\leq t-\Delta_x,t)\ln{\frac{\Delta_y\lambda_X(t_x\leq t-\Delta_x,t)}{c}}dt\nonumber\\
&=(\Delta_y)^{-1}\left[(1-c)\ln(1-c)+c(1+\ln(\Delta_y/\Delta_x))\right].
\label{stM}
\end{align}
Here we see an additional term as compared to the non-stationary result such that $\dot{M}^{\rm st}_X=\mathring{M}_X+(c/\Delta_y)\ln(\Delta_y/\Delta_x)$. This extra contribution over $\dot{M}_X$ arises from the ability of the full dynamics in $X$ to distinguish whether the system was in the spiking window, $[n\Delta_y,n\Delta_y+\Delta_x]$ or not over the time homogeneous Markov marginalisation process, characterised by $\lambda_X^{{\rm st},0}$, which both cannot detect the refractory period \emph{or} the existence of this window. Since $\lim_{t_0\to-\infty}\lambda_X[x_{[t_0,t)}]=\lambda(t_x,t)$, by definition, $\dot{T}_{\mathfrak{T}\to X}=0$, thus illustrating the specific case $\dot{M}_X^{\rm st}\geq\mathring{M}_X$ emerging from Eq.~(\ref{MgtM}) as claimed in Section~\ref{appA}.
\par
As we have seen, the process described in Section \ref{simple} leads to a disparity between $\dot{M}^{\rm st}_X$ and $\mathring{M}_X$ due to the time inhomogeneous spike rate in $X$, corresponding, in the framework described here, to $p_{\Phi}(\phi)=\delta(\phi)$. However, we can consider simple alterations to this process which change this property. Importantly, Eq.~(\ref{limlam}) holds for any $p_{\Phi}(\phi)$ and thus so does the final relation in Eq.~(\ref{lamst}) with the only exception being that $\phi$ need not equal $0$, but that corresponding to whatever value of $\phi$ is drawn from $p_{\Phi}(\phi)$. As such if we consider a process, identical to that in Section \ref{simple}, but where $y_{[t_0,t)}$ is generated such that it equals $y^*_{[\tau+\phi,t+\phi)}$ with probability density $p_{\Phi}(\phi)$ it will have $\dot{M}^{\rm st}_X$ equal to that in Eq.~(\ref{stM}) independently of $p_{\Phi}(\phi)$. Consequently, if we choose $p_{\Phi}(\phi)=\Delta_y^{-1}$, such that $Y$, and thus $X$ are both stationary we will have $\mathring{M}_X=\dot{M}_X=\dot{M}^{\rm st}_X$, again with $\dot{M}_X^{\rm st}$ given by Eq.~(\ref{stM}).
\par
Indeed we can give an expression for $\dot{M}_X$ for arbitrary  $p_{\Phi}(\phi)$ other than the $p_{\Phi}(\phi)=\Delta_y^{-1}$ and $p_{\Phi}(\phi)=\delta(\phi)$ we have already considered. To do this we first extend the domain of $p_{\Phi}(\phi)$ to $\phi\in\mathbb{R}$ such that it is periodic, i.e.  $p_{\Phi}(\phi+n\Delta_y)$ with $n\in\mathbb{Z}$, but retaining normalisation on $[0,\Delta_y)$ (i.e. $\int_0^{\Delta_y}p_{\Phi}(\phi')d\phi'=\int_\phi^{\phi+\Delta_y}p_{\Phi}(\phi')d\phi'=1$). We can then write an expression for $\lambda^0_X(t)$ as 
\begin{align}
\lambda^0_X(t)&=\frac{c}{\Delta_x}\int_{t-\Delta_x}^tp_{\Phi}(\phi)d\phi
\end{align}
thus expressing the difference $\dot{M}^{\rm st}_X-\mathring{M}_X$ as
\begin{align}
&\dot{M}^{\rm st}_X-\mathring{M}_X\nonumber\\
&=\int_0^{\Delta_y}\left[\frac{1}{\Delta_y}\int_{\phi}^{\phi+\Delta_y}e^{-\int_{\phi}^t \lambda_X(t_x\leq t-\Delta_x,t')dt'}\right.\nonumber\\
&\qquad\qquad\left.\times\lambda_X(t_x\leq t-\Delta_x,t)\ln\frac{\lambda^0_X(t)}{\lambda^{{\rm st},0}_X}dt\right]p_{\Phi}(\phi)d\phi\nonumber\\
&\geq 0,
\label{KLM}
\end{align}
where we have noted a lower bound due to its expression as a KL divergence, or more particularly, its expression as $\dot{T}^{(0)}_{\mathfrak{T}\to X}$, due to the fact $\dot{T}_{\mathfrak{T}\to X}=0$ as per Eq.~(\ref{MgtM}).
\par
 Continuing, we have,
\begin{align}
&\dot{M}^{\rm st}_X-\mathring{M}_X\nonumber\\
&=\int_0^{\Delta_y}\left[\frac{1}{\Delta_y}\int_{\phi}^{\phi+\Delta_x}\frac{c}{\Delta_x}\ln\frac{\lambda^0_X(t)}{\lambda^{{\rm st},0}_X}dt\right]p_{\Phi}(\phi)d\phi\nonumber\\
&=\frac{1}{\Delta_y}\int_0^{\Delta_y}f(\phi)p_{\Phi}(\phi)d\phi\nonumber\\
f(\phi)&=\int_{\phi}^{\phi+\Delta_x}\frac{c}{\Delta_x}\ln\left[\frac{\Delta_y}{\Delta_x}\int_{t-\Delta_x}^tp_{\Phi}(\phi')d\phi'\right]dt
\end{align}
such that
\begin{align}
\dot{M}^{\rm st}_X-\mathring{M}_X&=\frac{c}{\Delta_y}\ln\frac{\Delta_y}{\Delta_x}+\xi,
\end{align}
with
\begin{align}
\xi&=\frac{c}{\Delta_y\Delta_x}\int_{0}^{\Delta_y}p_{\Phi}(\phi)\int_{\phi}^{\phi+\Delta_x}\ln\left[\int_{t-\Delta_x}^tp_{\Phi}(\phi')d\phi'\right]dtd\phi.
\end{align}
The contents of the logarithm in $\xi$ always lies in $[0,1]$, due to the assertion $2\Delta_x<\Delta_y$ in the construction of the model, and so we have  $\xi\leq 0$. $\xi$ takes a maximum value $0$ when $p_{\Phi}(\phi)=\delta(\phi-a)$, $a\in[0,\Delta_y)$, with $a=0$ corresponding to the usage in Section \ref{simple}. Due to Eq.~(\ref{KLM}) being a KL divergence, it therefore takes a minimum value $-\frac{c}{\Delta_y}\ln\frac{\Delta_y}{\Delta_x}$ corresponding to $p_{\Phi}(\phi)=\Delta_y^{-1}$, i.e. when the process is stationary, such that 
\begin{align}
-\frac{c}{\Delta_y}\ln\frac{\Delta_y}{\Delta_x}\leq\xi\leq 0,
\end{align}
and in turn
\begin{align}
0\leq \dot{M}^{\rm st}_X-\mathring{M}_X\leq \frac{c}{\Delta_y}\ln\frac{\Delta_y}{\Delta_x}.
\end{align}
It is worth pointing out that such a process, despite being stationary when $p_{\Phi}(\phi)=\Delta_y^{-1}$, would not be ergodic for any choice of $p_{\Phi}(\phi)$ since once $\phi$ is drawn from the distribution, the process deterministically spikes with period $\Delta_y$ indefinitely.
\par
Finally, if only one sample, $\{x_{[\tau,t)},y_{[\tau,t)}\}$, is drawn, however long, from which empirical estimates are to be formed, then there is fundamental ambiguity as to the statistical nature of $Y$ and thus how large the over, or under, estimation of the active memory utilisation rate, $\dot{M}_X^{\rm st}-\mathring{M}_X$, is. \emph{If} it can be asserted that $Y$ is indeed drawn from a distribution $p_{\Phi}(\phi)$, only then may we state that it is an over-estimation that lies in $[0,\frac{c}{\Delta_y}\ln\frac{\Delta_y}{\Delta_x}]$ as per the above.
\end{document}